\documentclass[3p,twocolumn,number,sort&compress]{elsarticle}

\usepackage{tikz}
\usepackage{enumitem}

\usepackage{graphicx}  % Include figure files
\usepackage{multirow}

\usepackage{fancyhdr}
\usepackage{float,placeins}
\usepackage{longtable}
\usepackage{parskip}
\usepackage[T1]{fontenc}
\usepackage{dcolumn}   % Align table columns on decimal point
\usepackage{bm}        % bold math
\usepackage{amsfonts}  % Common math fonts
\usepackage{amsmath}   % Common math functions
\usepackage{amssymb}   % Common math symbols
\usepackage{xcolor}
\usepackage{bbm}
\usepackage{url,hyperref}
\hypersetup{colorlinks=true, linkcolor=BrickRed,
  urlcolor=blue!50!black, citecolor=blue!50!black}
\usepackage[capitalize]{cleveref}
\let\ref\cref
\usepackage{xpatch}
\usepackage{appendix}
\crefrangelabelformat{equation}{(#3#1#4)$-$(#5#2#6)}
\xapptocmd\appendices{%
  \crefalias{section}{appendix}%
}{}{\PatchFailed}

\makeatletter
\newcommand{\dbloverline}[1]{\overline{\dbl@overline{#1}}}
\newcommand{\dbl@overline}[1]{\mathpalette\dbl@@overline{#1}}
\newcommand{\dbl@@overline}[2]{%
  \begingroup
  \sbox\z@{$\m@th#1\overline{#2}$}%
  \ht\z@=\dimexpr\ht\z@-2\dbl@adjust{#1}\relax
  \box\z@
  \ifx#1\scriptstyle\kern-\scriptspace\else
  \ifx#1\scriptscriptstyle\kern-\scriptspace\fi\fi
  \endgroup
}
\newcommand{\dblunderline}[1]{\@@underline{\dbl@underline{#1}}}
\newcommand{\dbl@underline}[1]{\mathpalette\dbl@@underline{#1}}
\newcommand{\dbl@@underline}[2]{%
  \begingroup
  \sbox\z@{$\m@th#1\@@underline{#2}$}%
  \dp\z@=\dimexpr\dp\z@-2\dbl@adjust{#1}\relax
  \box\z@
  \ifx#1\scriptstyle\kern-\scriptspace\else
  \ifx#1\scriptscriptstyle\kern-\scriptspace\fi\fi
  \endgroup
}
\newcommand{\dbl@adjust}[1]{%
  \fontdimen8
  \ifx#1\displaystyle\textfont\else
  \ifx#1\textstyle\textfont\else
  \ifx#1\scriptstyle\scriptfont\else
  \scriptscriptfont\fi\fi\fi 3
}
\makeatother

\newcommand{\mat}[1]{{%
  \mspace{0.5mu}%
  \dblunderline{\mspace{-0.5mu}#1_{}\kern-\scriptspace\mspace{-0.5mu}}%
  \mspace{0.5mu}%
  \mathcorr{#1}%
}}
\makeatletter
\newcommand{\mathcorr}[1]{\mathpalette\math@corr{#1}}
\newcommand{\math@corr}[2]{%
  \begingroup
  \sbox\z@{$\m@th#1#2$}\sbox2{$\m@th#1#2_{}\kern-\scriptspace$}%
  \kern\dimexpr\wd\z@-\wd\tw@\relax
  \endgroup
}
\makeatother

\graphicspath{{Figures/}}

\setcitestyle{super}

\newcommand{\pwisein}{\left\{ \begin{array}{ll}}
\newcommand{\pwiseout}{\end{array}\right.}

\newcommand{\complexi}{\mathbbm{i}}
\renewcommand{\vec}[1]{\mathbf{#1}}

\begin{document}

\begin{frontmatter}
\title{Independent Control of Transport and Order in a Ratcheted Colloidal Suspension}

\author[1,fn1]{Sudipta Mandal}
\ead{ph16049@iisermohali.ac.in}
\ead{sudiptam@tifrh.res.in}
\author[1]{Dipanjan Chakraborty \corref{cor1}}
\ead{chakraborty@iisermohali.ac.in}
\author[2,3]{Debasish Chaudhuri }
\ead{debc@iopb.ac.in}

\affiliation[1]{organization={Department of Physical Science, Indian Institute of
    Science Education and Research Mohali},
  addressline={Sector 81, S. A. S
    Nagar,Manauli PO 140306,India}}

\affiliation[2]{organization={Institute of Physics},
  addressline={P.O: Sainik School,
  Bhubaneswar-751005, Odisha, India}}

  \affiliation[3]{organization={Homi Bhabha National Institute},
  addressline={Training School Complex, Anushakti Nagar, Mumbai-400094, India }}
\fntext[fn1]{Present affiliation:Tata institute of fundamental research,Hyderabad
36/P, Gopanpally Village, Serilingampally Mandal, Ranga Reddy District, Hyderabad, Telangana 500046}
\cortext[cor1]{Corresponding author}

\date{\today}

\begin{abstract}  

  We study directed transport in a two-dimensional suspension of
  repulsively interacting colloids driven by a stochastic asymmetric
  piecewise-linear flashing ratchet using large-scale molecular
  dynamics simulations. The driving frequency and the ratchet
  asymmetry offer two independent ways of controlling the particle
  current, but they affect the suspension differently. At fixed
  asymmetry, the current shows a resonance with ratcheting frequency
  that is set by the collective relaxation dynamics of the interacting
  particles. The resulting increase in transport is accompanied by
  defect-mediated structural changes, showing density-dependent
  hexatic and solid-like states, with larger currents generally
  associated with weaker ordering. By contrast, at fixed frequency,
  changing the ratchet asymmetry mainly alters the strength of the
  directed bias and can significantly enhance the current while
  leaving the hexatic order largely unchanged. Near the equilibrium
  hexatic-melting regime, this makes it possible to generate
  substantial directed currents without strongly disrupting sixfold
  orientational order. These results show that frequency tuning
  couples transport to structural reorganization, whereas asymmetry
  tuning primarily controls transport leaving the structure largely
  unaltered, providing distinct and complementary routes for
  manipulating transport and order in driven colloidal suspensions.

\end{abstract}
\end{frontmatter}
%For fun, see PACS list at  https://www.aip.org/publishing/pacs/pacs-2010-regular-edition
%\pacs{47.15.-x}

%\maketitle 

\section{Introduction}

Non-equilibrium driven systems, or pump models, can generate a net
directed current even when the applied forces vanish under
spatio-temporal averaging~\cite{Julicher1997a, Astumian2002,
  Hanggi2009, Reimann2002, Citro2003, Chaudhuri2011, Chaudhuri2015,
  Chaudhuri2015f, Cubero2016}. Such transport requires broken detailed
balance together with spatial or temporal asymmetry, and underlies
many biological processes, including ion pumps and molecular
motors~\cite{Gadsby2009, Reimann2002,Molcrette2022}. A minimal theoretical framework
is provided by flashing ratchet models~\cite{Julicher1997a}, which
have also been realized experimentally in colloidal suspensions using
optical~\cite{Faucheux1995, Wei1998, Lopez2008,Diwakar2024,tang2021,Wen2025},
magnetic~\cite{Tierno2010, Tierno2012}, and electric field–based
protocols~\cite{Rousselet1994, Leibler1994, Marquet2002,Herman2023}.

Most studies of ratchet transport have focused on non-interacting
particles, where directed motion arises primarily from single-particle
rectification. In contrast, dense interacting colloidal suspensions
offer a qualitatively different setting in which transport is
inherently coupled to collective relaxation, structural order, and
defect dynamics. This makes them a particularly useful platform for
understanding how directed motion can emerge without necessarily
disrupting local crystalline order. Earlier work has explored aspects
of interaction-induced collective transport and structural response in
related pump models~\cite{Derenyi1995, Derenyi1996, Chakraborty2014,
  rossini2018sliding, martin2025colloidal, Jain2007, Marathe2008,
  Chaudhuri2011, Chaudhuri2015, Chaudhuri2015f, patil2022using,
  Savelev2004, Pototsky2010, Savelev2003, Hanggi2009}, including in
two-dimensional colloidal systems driven by time-dependent potentials.

In our earlier work~\cite{khali2020structure}, we studied a
two-dimensional colloidal suspension driven by a stochastically
switching asymmetric periodic potential at fixed asymmetry. That study
revealed density- and frequency-dependent solid, hexatic, and
re-entrant solid–hexatic–solid behavior, with maximal current
occurring in a dynamically disordered regime. However, how directed
transport and nonequilibrium phases respond when the spatial asymmetry
of the ratchet potential itself is varied remained an open question.

In this work, we address this issue by considering a two-dimensional
suspension of repulsively interacting colloids subjected to a
stochastic asymmetric piecewise-linear flashing ratchet that generates
directed motion in an otherwise unbiased environment. The system
combines collective many-body dynamics with a time-dependent
nonequilibrium drive, providing a minimal setting for exploring the
interplay between transport, structural organization, and defect
formation in driven soft matter. To characterize this interplay, we
perform large-scale molecular dynamics simulations over a broad range
of densities, ratcheting frequencies, and potential asymmetries,
allowing us to probe both the transport response and the accompanying
structural evolution of the suspension.

Our central result is that frequency tuning and asymmetry tuning
control directed transport in fundamentally different ways: the former
couples transport to defect-mediated structural reorganization,
whereas the latter primarily regulates transport while leaving the
structure largely unchanged. This distinction enables substantial
currents to be generated without sacrificing orientational order and
provides complementary strategies for controlling transport and
organization in driven colloidal matter.

The remainder of the paper is organized as follows. In
Sec.~\ref{sec:theory}, we introduce the model and numerical
methods. Sec.~\ref{sec_results} presents the main results, including
the dependence of the ratcheted current and structural properties on
control parameters, the interplay between transport and structure,
defect formation, and the resulting nonequilibrium phase
diagram. Finally, Sec.~\ref{sec_outlook} provides a summary and
outlook.

\section{Model and Simulation details}
\label{sec:theory}

We consider a two-dimensional suspension of $N$ repulsively
interacting colloids in an area $A=L_xL_y$, with number density
$\rho=N/A$. At high enough densities, the equilibrium system displays
a solid phase~\cite{Kapfer:2015ca}. The mean triangular-lattice
spacing is defined by $a^2=2/(\sqrt{3}\rho)$, and the separation
between neighbouring lattice planes is $a_y=\sqrt{3}a/2$. The
particles interact through a truncated soft repulsion
\begin{equation}
  U(r)=\epsilon\left[\left(\frac{\sigma}{r}\right)^{12}-2^{-12}\right],
  \qquad r<r_c,
\end{equation}
with cutoff $r_c=2\sigma$. The energy and length scales are set by
$\epsilon$ and $\sigma$, and the time scale is
$\tau=\sqrt{m\sigma^2/\epsilon}$. The system is maintained at
$k_{\rm B}T/\epsilon=1$ using a Langevin thermostat with friction
coefficient $\gamma=1/\tau$. In equilibrium, the corresponding
soft-core solid undergoes a two-step solid--hexatic--liquid melting
transition with the solid melting density close to
$\rho_c\sigma^2\simeq 1.01$~\cite{Kapfer:2015ca}.

The suspension is driven along a direction perpendicular to one set of
lattice planes, say the $y$ axis, by a commensurate asymmetric
flashing ratchet periodic potential,
\begin{equation}
  U_{\rm ext}(y,t)=s(t)u(y), \qquad u(y+\lambda)=u(y),
\end{equation}
where the spatial period is chosen as $\lambda=a_y$. The stochastic variable
$s(t)$ switches between $0$ and $1$ with equal on -- off and off -- on
rates, so that
$p_{{\rm on}\to{\rm off}}=p_{{\rm off}\to{\rm on}}$. The potential
profile within one period in the on-state is
\begin{equation}
 u(y) =
 \begin{cases}
 (2U_0/\lambda)(1-\delta)^{-1} y,
 & 0 \le y \le y_0, \\
 (2U_0/\lambda)(1+\delta)^{-1}(\lambda-y),
 & y_0 \le y \le \lambda,
 \end{cases}
 \label{eq:external_potential_1}
\end{equation}
with $y_0=(1-\delta)\lambda/2$ and $U_0=\epsilon$. The parameter
\begin{equation}
  \delta=\frac{s_2-s_1}{s_2+s_1}
\end{equation}
measures the asymmetry of the two segments of the potential. Unlike
the ratio $\alpha=s_1/s_2$, the parameter $\delta$ is bounded between
$-1$ and $1$, and reversing the potential asymmetry direction
corresponds simply to $\delta\to -\delta$.  Therefore, under this
transformation, the direction of the current is expected to reverse,
while its amplitude is expected to be proportional to $\delta$.

Molecular dynamics simulations are performed using the leap-frog
algorithm~\cite{frenkel2002algorithms} with time step
$\delta t=0.001\tau$. We set $\epsilon=\sigma=m=1$ in the simulations.
After equilibration, stochastic ratcheting is implemented at each time
step by attempting to switch $s(t)$ between $0$ and $1$ with
probability $f\delta t$. We present results for $N=65536$ particles.
The first $10^7$ steps are discarded to reach the steady state, and
measurements are collected over a further $10^7$ steps.

\section{Results}
\label{sec_results}

Let us first emphasize the central question of this study: how can one
generate directed transport in a driven colloidal suspension without
necessarily destroying its underlying structure?  Notably, the ratchet
mechanism introduces two independent external control parameters.  The
driving frequency controls how the suspension explores the periodic
potential in time, and is therefore expected to couple to the
collective relaxation of the particles. The asymmetry parameter, on
the other hand, controls the spatial bias of the potential. We show
below that these two controls generate current through different
physical routes. Frequency tuning changes both transport and
structure, whereas asymmetry tuning primarily changes the magnitude of
the current.

\subsection{Current in Interacting System}
\label{sec:current_interacting}

We first examine the time-averaged directed particle current as a
function of the two control parameters: frequency tuning and asymmetry
tuning. In the dilute regime, interactions only weakly perturb the
single-particle ratchet response, so the current density increases
with particle density. At higher densities, however, interactions
become dominant. Collisions, caging, and emerging structural
correlations hinder motion across the potential landscape, causing the
current to decrease. The resulting nonmonotonic density dependence
signals a crossover from a weakly interacting ratchet regime to
transport limited by collective relaxation.

\begin{figure*}[!t]
  \centering
  \includegraphics[width=0.9\textwidth]{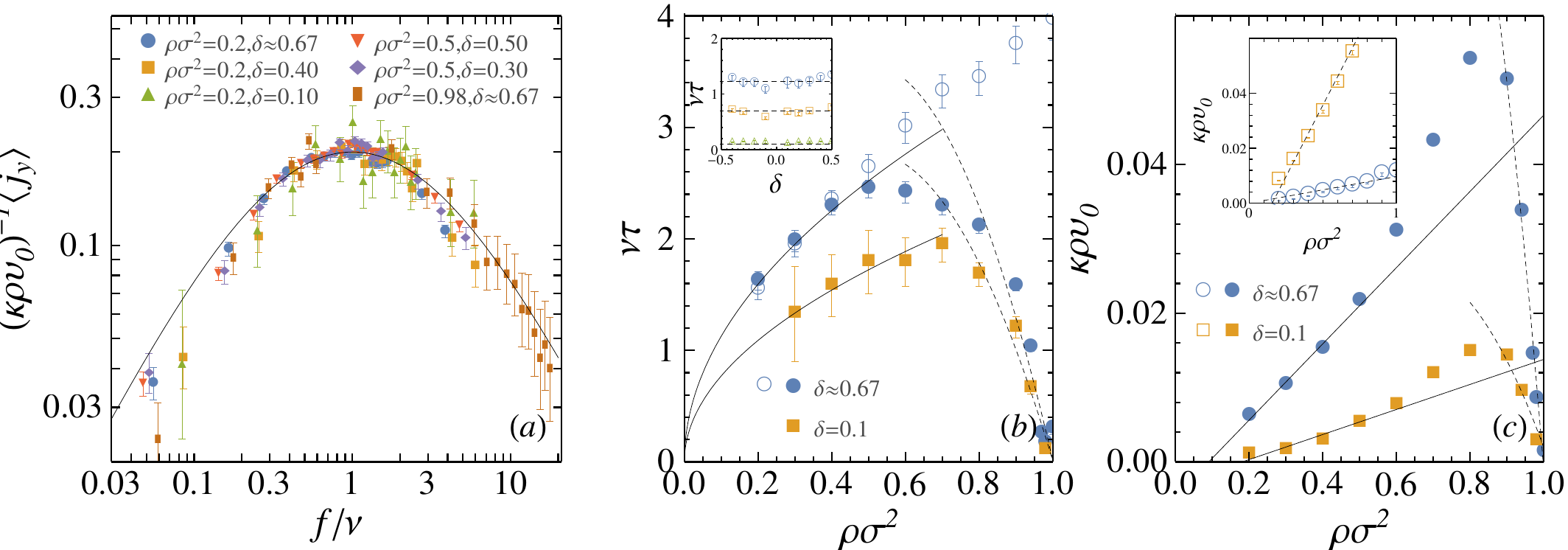}
  \caption{ \textbf{(a)} Collapse of the dimensionless average current
    $(\kappa\rho v_0)^{-1}\langle j_y\rangle$ as a function of the
    scaled frequency $f/\nu$ for the densities and asymmetry
    parameters indicated in the legend. The solid line shows the
    interpolation function given by \cref{eq:g_nu_f}. \textbf{(b)}
    Resonance frequency $\nu$, obtained by fitting the current data
    using \cref{eq:current_anstaz,eq:g_nu_f}, as a function of
    density. Open circles denote the non-interacting result, for which
    $\nu$ increases monotonically with density. Solid lines are fits
    based on $\nu \sim \rho^{1/2}$, characteristic of ballistic
    transport, while dashed lines are fits based on
    $\nu \sim \rho(1-\rho/\rho_c)$, characteristic of crowded
    diffusive transport. (Inset) The relaxation rate $\nu$ remains
    nearly independent of the asymmetry parameter, as
    expected. \textbf{(c)} Current amplitude $j_0=\kappa\rho v_0$ as a
    function of density. The amplitude is non-monotonic, reaching a
    maximum near $\rho\sigma^2 \approx 0.85$. At low densities, $j_0$
    grows linearly with $\rho$, consistent with ballistic transport
    (inset; free particle results). With increasing density, it
    crosses over to $j_0 \sim \rho^{3/2}(1-\rho/\rho_c)$, as indicated
    by the dashed lines. The corresponding fit yields
    $\rho_c\sigma^2 \approx 1.01$, close to the equilibrium melting
    density.  }
  \label{fig:current_dens_int_nint}
\end{figure*}

The time-averaged directed current along the drive is expressed as
\begin{equation}
  \langle j_y\rangle=
  \frac{1}{\tau_m L_xL_y}
  \int_0^{\tau_m}dt\int_0^{L_y}dy\int_0^{L_x}dx\, j_y(x,y,t),
\end{equation}
where $\tau_m$ is chosen as an integer multiple of the mean switching
time, $1/f$, of the ratchet. 

For fixed density and asymmetry, the current is suppressed in both the
low- and high-frequency limits. At low frequencies, particles
equilibrate within each potential state, while at high frequencies the
potential switches too rapidly for particles to respond. Consequently,
the current exhibits a maximum at an intermediate frequency comparable
to the intrinsic relaxation rate of the suspension.

Based on the above physical picture, we introduce the following ansatz
for the current~\cite{Chakraborty2014,khali2020structure}
\begin{equation}
  \langle j_y\rangle=j_0 g(\nu,f),
  \label{eq:current_anstaz}
\end{equation}
where $j_0=\kappa\rho v_0$ is the current amplitude and $\nu$ is the intrinsic 
relaxation frequency of the colloidal dispersion. The interpolation form
\begin{equation}
  \label{eq:g_nu_f}
  g(\nu,f)=\frac{\nu f}{\nu^2+c\nu f+f^2}
\end{equation}
recovers the observed low-frequency growth and high-frequency decay, and
has a maximum at $f=\nu$.

The current data are fitted to \cref{eq:current_anstaz,eq:g_nu_f}
using three parameters: the amplitude $j_0$, the resonance frequency
$\nu$, and a dimensionless constant $c$. Across the parameter range
studied, the fits yield $c \simeq 3$. With this value, the rescaled
data collapse onto a single master curve
(\cref{fig:current_dens_int_nint}(a))~\cite{Mandal2025u}. The density
dependence of the current is captured by two quantities: the
relaxation frequency $\nu$ and the amplitude $j_0$.

The density dependence of the relaxation frequency reflects the
underlying transport mechanism. In the dilute regime, the relevant
timescale is the ballistic traversal time across one period of the
potential. Since the potential period is commensurate with the mean
interparticle spacing, $\lambda^2 \sim \rho^{-1}$. Combining this with
the kinematic relation between force and acceleration yields
$\nu \sim (U_0 \rho)^{1/2}$, and the speed $v_0 \sim U_0$. The
corresponding current is
\begin{equation}
  \label{eq:current_low_density}
  \langle j_y\rangle=
  \kappa\rho^{3/2}
  \frac{fU_0}{f^2+3f(\rho U_0)^{1/2}+\rho U_0}.
\end{equation}
At high density, the motion across one period is no longer ballistic-like
but diffusion-limited. Using
$\nu\sim D(\rho)/\lambda^2=\rho D(\rho)$,
$v_0\sim \rho^{1/2}D(\rho)$, and
$D(\rho)=D_0(1-\rho/\rho_c)$~\cite{khali2020structure}, we obtain
\begin{equation}
    \langle j_y \rangle =
    \kappa
    \frac{fD_0^2\rho^{5/2}(1-\rho/\rho_c)^2}
    {D_0^2\rho^2(1-\rho/\rho_c)^2
      +3fD_0\rho(1-\rho/\rho_c)+f^2}.
    \label{eq:current_expression}
\end{equation}

Figure \cref{fig:current_dens_int_nint}(b) shows the density
dependence of the relaxation rate $\nu$. For interacting particles,
$\nu$ exhibits a crossover from the ballistic scaling
$\nu \sim \rho^{1/2}$ at low density to the diffusive form
$\nu \sim \rho D_0(1-\rho/\rho_c)$ at higher density, as indicated by
the solid and dashed lines, respectively. The suppression of $D(\rho)$
near $\rho_c$ reduces the relaxation rate. The fitted value
$\rho_c\sigma^2 \simeq 1.01$ is close to the equilibrium melting
density, indicating a high density dynamical arrest.

Figure \cref{fig:current_dens_int_nint}(c) shows the corresponding
current amplitude $j_0$. At low density, the data are consistent with
the ballistic scaling $j_0 \sim \rho U_0$ (solid lines), whereas at
higher density they follow the diffusive prediction
$j_0 \sim \rho^{3/2}(1-\rho/\rho_c)$ (dashed lines). The decrease of
$D(\rho)$ at high density suppresses the current amplitude and leads
to a maximum at intermediate density, indicating that optimal
transport occurs close to the ballistic-to-diffusive crossover.

\begin{figure}[!t]
  \centering
  \includegraphics[width=\linewidth]{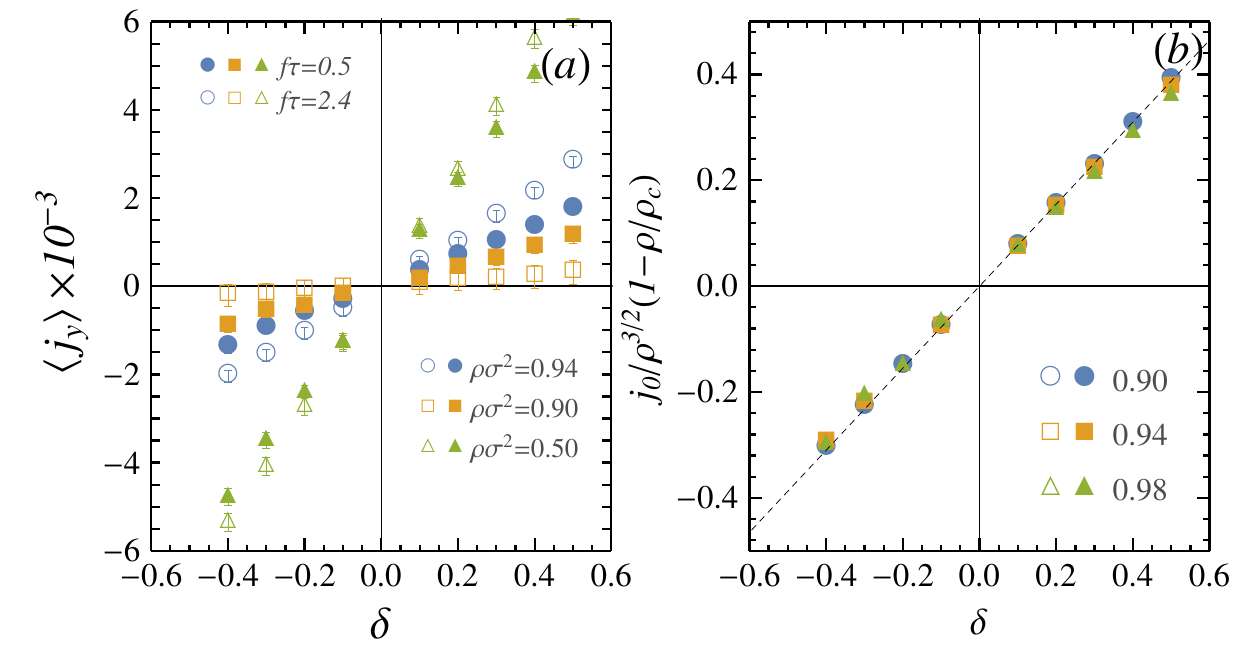}
  \caption{\textbf{(a)} Time-averaged current as a function of the
    asymmetry parameter for three densities and two driving
    frequencies. At fixed density, the curves do not collapse because
    of the frequency dependence of $g(\nu,f)$ and $j_0$. \textbf{(b)}
    Scaled amplitude $\kappa=j_0/[\rho^{3/2}D_0(1-\rho/\rho_c)]$
    versus $\delta$. The data collapse onto a linear master curve
    (dashed line), demonstrating proportionality to the potential
    asymmetry.  }
  \label{fig:current_data}
\end{figure}

{\it Impact of asymmetry parameter.}  Figure
\cref{fig:current_data}(a) shows the directed current as a function of
the asymmetry parameter $\delta$ at fixed density and driving
frequency. The current increases approximately linearly with $\delta$,
suggesting that the spatial asymmetry primarily controls the magnitude
and direction of transport. Since $\langle j_y\rangle$ also depends on
the frequency response $g(\nu,f)$ and the density-dependent amplitude
$j_0$, however, the raw current alone does not directly reveal the
underlying dependence on $\delta$.

To isolate this dependence, we extract $j_0$ from the fits and define
\begin{equation}
\kappa=\frac{j_0}{\rho^{3/2}D_0(1-\rho/\rho_c)}.
\end{equation}
The resulting values of $\kappa$ collapse onto a single line for different densities, as shown in \cref{fig:current_data}(b), yielding
\begin{equation}
\kappa=m\delta,
\end{equation}
with $m\simeq 0.8$. This collapse shows that the asymmetry acts as a
geometric bias that sets the strength and sign of the directed
current, while having little effect on the resonance
frequency. Frequency and asymmetry therefore play complementary roles:
frequency determines the collective dynamical response, whereas
asymmetry biases that response to generate directed transport.

\subsection{Order and Particle Current}
\label{sec:order_current}

Another key question is whether the currents generated by the two
protocols carry the same structural cost. At $k_{\rm B}T/\epsilon=1$,
the equilibrium system melts in two stages, from a triangular solid to
an isotropic liquid through an intermediate hexatic phase. The solid
melts at $\rho_s\sigma^2 \approx 1.014$ to hexatic, while the hexatic
phase spans $0.997 \lesssim \rho\sigma^2 \lesssim 1.001$. A
commensurate periodic potential stabilizes solid and shifts these
transitions to lower densities, with solid-hexatic melting occurring
near $\rho\sigma^2 \approx 0.95$.~\cite{Chaudhuri2006} This places the
densities studied here close to the melting regime, allowing us to
test whether large ratchet currents can coexist with sixfold
orientational order.

\begin{figure}[!t]
  \centering
  \includegraphics[width=\linewidth]{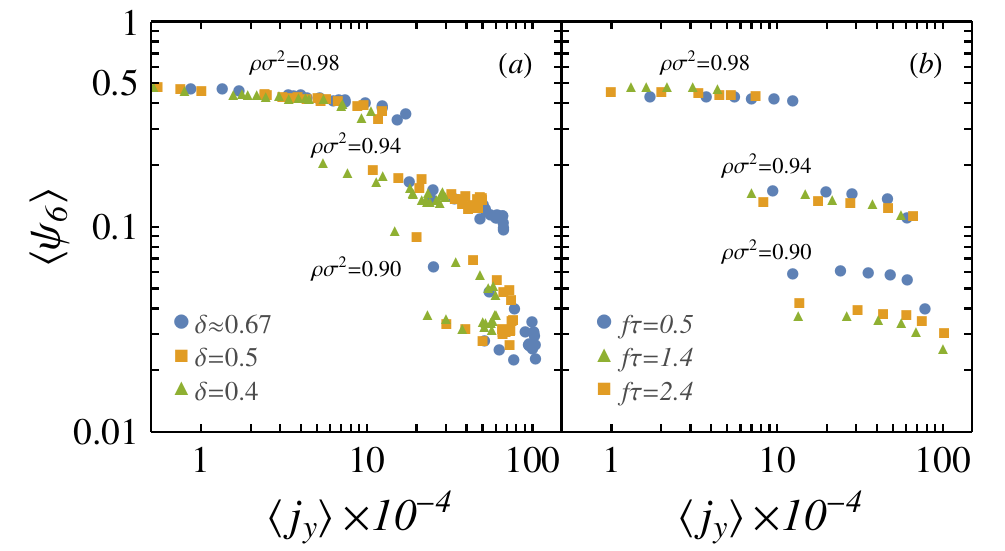}
  \caption{\textbf{(a)} Parametric plot of the hexatic order parameter
    $\langle \psi_6\rangle$ versus the particle current
    $\langle j_y\rangle$, with the driving frequency eliminated at
    fixed density and asymmetry parameter. \textbf{(b)} Corresponding
    plot with the asymmetry parameter eliminated at fixed density and
    driving frequency. Fixed parameter values are listed in the
    legends.  }
  \label{fig:order_particle_current}
\end{figure}

To quantify the global hexatic order, we use the order parameter defined as
\begin{equation}
  \label{eq:global_psi6}
   \left \langle \psi_6 \right \rangle
   =
   \left \langle
   \left |
   \frac{1}{N} \sum_{i=1}^{N}\psi_6^i
   \right |^2
   \right \rangle ,
\end{equation}
where
\begin{equation}
  \label{eq:local_psi6}
  \psi_6^i
  =
  \frac{1}{n_i}
  \sum_{k=1}^{n_i}
  e^{\complexi 6\theta_{ik}} .
\end{equation}
Here $n_i$ is the number of Voronoi neighbours of particle $i$, and
$\theta_{ik}$ is the angle made by the separation vector between
particles $i$ and $k$ with the $x$-axis.

To compare transport and structural order, we eliminate one control
parameter at a time. At fixed density and asymmetry, both
$\langle j_y\rangle$ and $\langle \psi_6\rangle$ depend on the driving
frequency. Eliminating $f$ yields the parametric curves in
\cref{fig:order_particle_current}(a). At fixed density and frequency,
eliminating the asymmetry parameter instead gives
\cref{fig:order_particle_current}(b).

When the frequency is eliminated at fixed density and asymmetry
parameter (\cref{fig:order_particle_current}(a)), the data separate by
density. Across densities, larger driven currents are consistently
associated with reduced hexatic order. This trend is evident from low
to high density: near the peak-current regime
($\rho\sigma^2 \approx 0.90$) the current is large and hexatic order
is weak, whereas at intermediate density ($\rho\sigma^2 \approx 0.94$)
the current is reduced as hexatic order increases, and close to the
hexatic regime ($\rho\sigma^2 \approx 0.98$) the current is further
suppressed while hexatic order is substantial.

However, this anticorrelation between current and hexatic order is
strongly suppressed when the current is increased via the asymmetry of
the ratcheting potential. In \cref{fig:order_particle_current}(b), we
show parametric plots of $\langle j_y\rangle$ versus
$\langle \psi_6\rangle$, obtained by eliminating the asymmetry
parameter at fixed density and driving frequency. Increasing $\delta$
produces a significant increase in $\langle j_y\rangle$ leaving
$\langle \psi_6\rangle$ nearly unchanged. Thus large currents can be
generated with only weak suppression of hexatic order, offering an
important practical consequence of controlling the asymmetry.

Asymmetry tuning provides a more selective control of transport than
frequency variation. At fixed density and driving frequency,
increasing $\delta$ enhances the directed current while leaving
$\langle \psi_6\rangle$ nearly unchanged, cleanly separating transport
from structure in the asymmetric ratchet. This contrasts with
frequency control, where changes in current are generally accompanied
by concomitant changes in the structural state.

This separation is most pronounced near the hexatic regime: frequency
variation couples enhanced transport to reduced hexatic order, whereas
asymmetry tuning enables substantial increases in current without
appreciably shifting $\langle \psi_6\rangle$.

Overall, asymmetry offers a route to boost directed currents without
significantly perturbing the underlying structure, in contrast to
frequency control, which intrinsically links transport to structural
rearrangements.

\subsection{Nonequilibrium Phase Behaviour}
\label{sec:noneq_phase_behavior}

We now proceed to analyze the structural order, under ratcheting
drive, in more detail. The hexatic order parameter
$\langle\psi_6\rangle$ measures sixfold orientational order, while the
solid order parameter $\langle\psi_G\rangle$ measures positional
order. It is defined as follows.  The structure factor of the system
is given by
\begin{equation}
    S({\bf q}) = \langle \rho_{\bf q} \rho_{-{\bf q}} \rangle
\end{equation}
where
$\rho_{\bf q} = \frac{1}{N} \sum_{i=1}^N \exp( i {\bf q} \cdot {\bf
  r}_i)$.  In a perfect triangular lattice, quasi-Bragg peaks emerge
at $ {\bf q}_p = (0, \pm 2\pi/a_y), (\pm 2\pi/a, \pm \pi/a_y) $.  The
solid order parameter $ \langle \psi_{\bf G} \rangle $ is calculated
by taking an average over $ S({\bf q}) $ values calculated at the six
quasi-Bragg peaks located at $ {\bf G}:= \{ {\bf q}_p \} $:
\begin{equation}
  \langle \psi_{\bf G} \rangle = \frac{1}{6}\sum_{p=1}^6 \left\langle  \left | \frac{1}{N} \sum_{i=1}^N \exp( i {\bf q}_p \cdot {\bf r}_i) \right | \right\rangle
\end{equation}
Since the periodic ratchet potential in the on-state directly induces
a density modulation along the $y$ direction, we compute the solid
order using only the subset of reciprocal lattice vectors
$\mathbf{G}_2 = (\pm 2\pi / a, \pm \pi / a_y)$, which isolates the
contributions relevant to spontaneous symmetry breaking (see
\cref{appendix:equilibrium_melting} for identification of the solid
and hexatic phase in equilibrium melting).  A driven state can lose
positional order while retaining orientational order.

\begin{figure}[!htbp]
\centering
\includegraphics[width=0.8\linewidth]{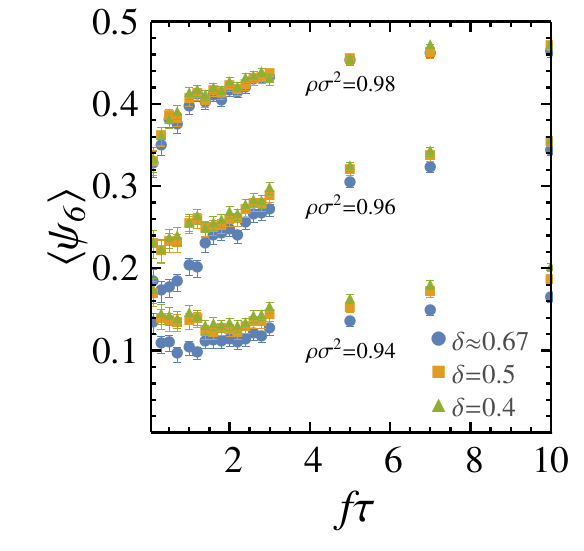}
\caption{Hexatic order parameter $\langle\psi_6\rangle$ versus driving
  frequency $f\tau$ for $\rho\sigma^2 = 0.94, 0.96,$ and $0.98$, and
  three values of the asymmetry parameter. At $\rho\sigma^2 = 0.98$,
  higher frequencies increases hexatic order more sharply than at
  smaller densities.  }
\label{fig:psi6_vs_freq}
\end{figure}

\begin{figure}[!htbp]
  \centering
  \includegraphics[width=\linewidth]{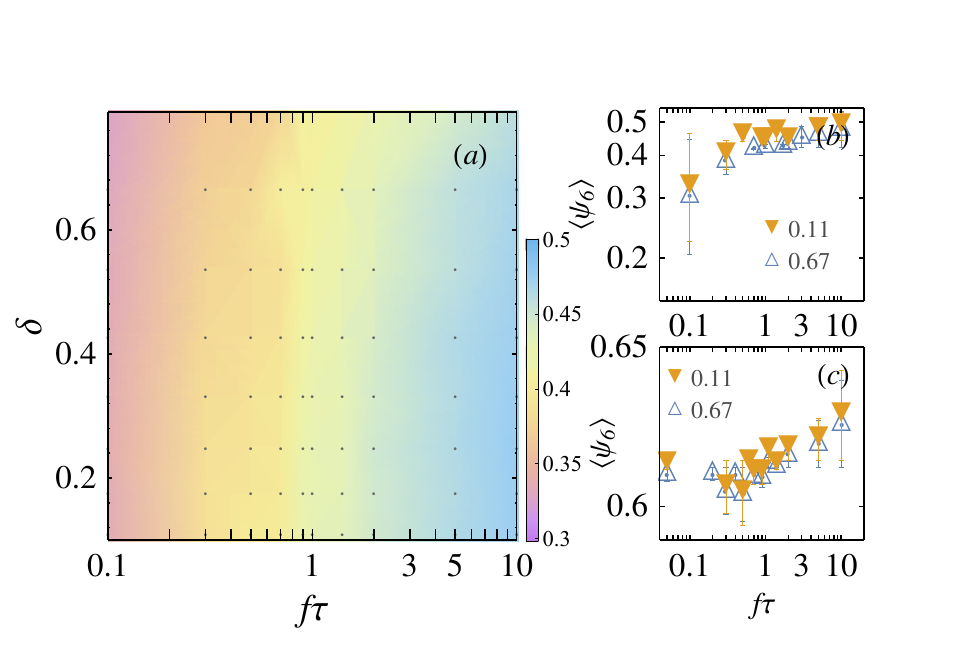}
  \caption{\textbf{(a)} Heat map of the hexatic order parameter
    $\langle\psi_6\rangle$ in the frequency–asymmetry plane at
    $\rho\sigma^2=0.98$. Symbols denote simulation state
    points. \textbf{(b)} Cuts at $\delta=0.11$ and $0.67$, showing a
    gradual increase of $\langle\psi_6\rangle$ with frequency and weak
    dependence on asymmetry. \textbf{(c)} Corresponding cuts at
    $\rho\sigma^2=1.04$, where orientational order is higher and
    remains nearly unchanged up to the resonance regime.}
  \label{fig:hexatic_density_plot}
\end{figure}

\begin{figure*}[!htbp]
  \centering
  \includegraphics[width=\linewidth]{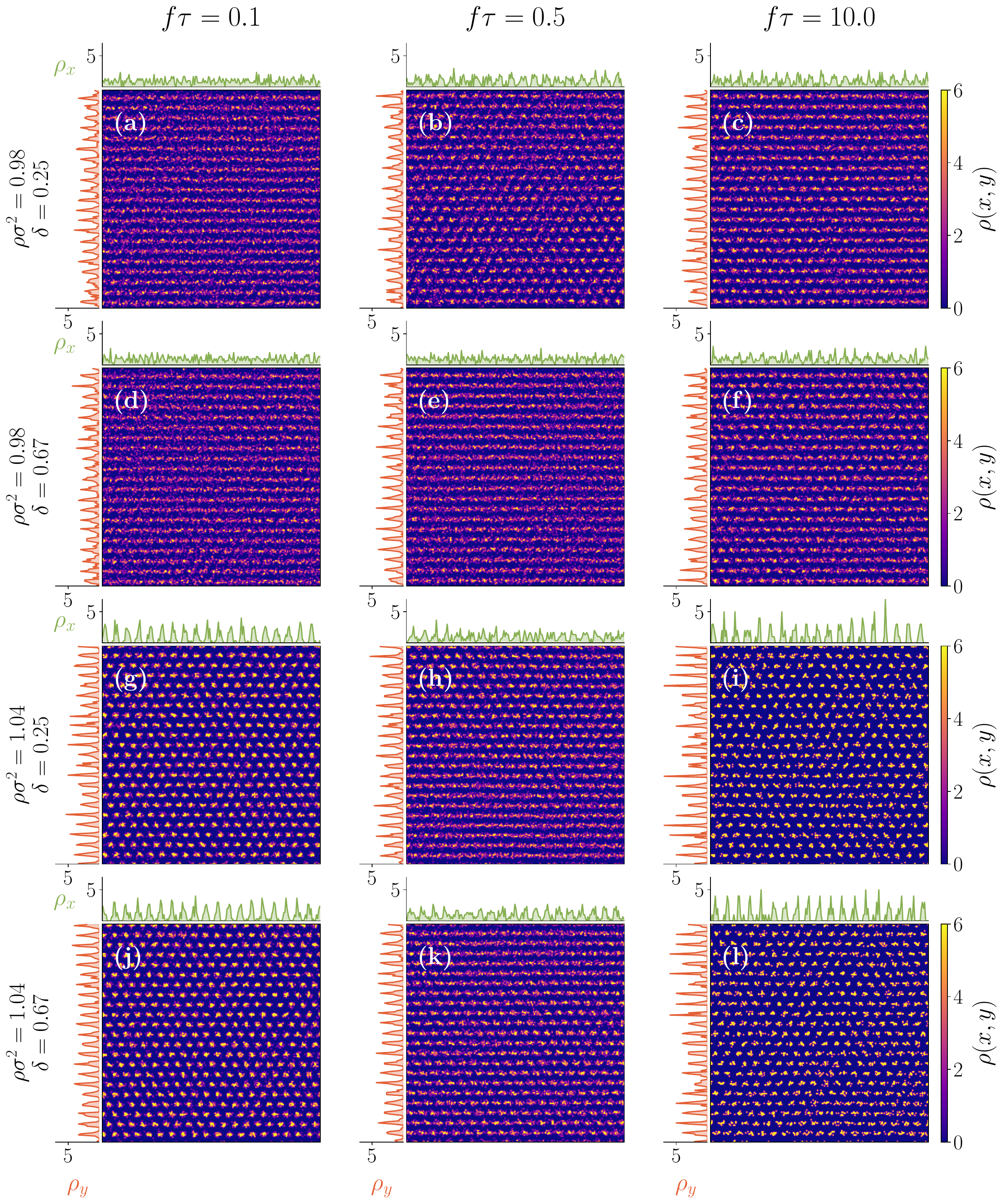}
  \caption{
    Time-averaged density profiles $\rho(x,y)$ for two densities,
    $\rho\sigma^2=0.98$ (figures (a) --(f)) and $1.04$ (figures (g)
    --(f)), two asymmetry parameters, $\delta=0.25$ and $0.67$, and
    three driving frequencies, $f\tau=0.1$, $0.5$, and $10$.  On top
    and the left side of every panel are shown the cuts of the density
    profile along the $x$-direction ($\rho_x$) and along the
    $y$-direction ($\rho_y$), respectively.  At $\rho\sigma^2=0.98$,
    the system shows mainly density modulation at low and intermediate
    frequencies, while a soft triangular pattern appears at high
    frequency. At $\rho\sigma^2=1.04$, a triangular structure is
    visible at low and high frequencies, but is weakened at
    intermediate frequency, indicating re-entrant melting. The time
    averaging was done over $50$ independent snapshots.}
  \label{fig:rhoxy}
\end{figure*}

\cref{fig:psi6_vs_freq} shows how $\langle\psi_6\rangle$ depends on
density and driving frequency. The data form three clear clusters
corresponding to $\rho\sigma^2=0.94$, $0.96$, and $0.98$. Near the
hexatic regime ($\rho\sigma^2=0.98$), $\langle\psi_6\rangle$ remains
appreciable over the full frequency range and increases weakly with
$f\tau$. At $\rho\sigma^2=0.96$, the magnitude is reduced but retains
a similar trend, while at $\rho\sigma^2=0.94$ the hexatic order is
weak and only weakly frequency dependent. In contrast, the three
values of $\delta$ collapse at each density, indicating that
$\langle\psi_6\rangle$ is governed primarily by density and frequency,
with only a weak dependence on the potential asymmetry.

\begin{figure*}[!ht]
\centering
\includegraphics[width=\linewidth]{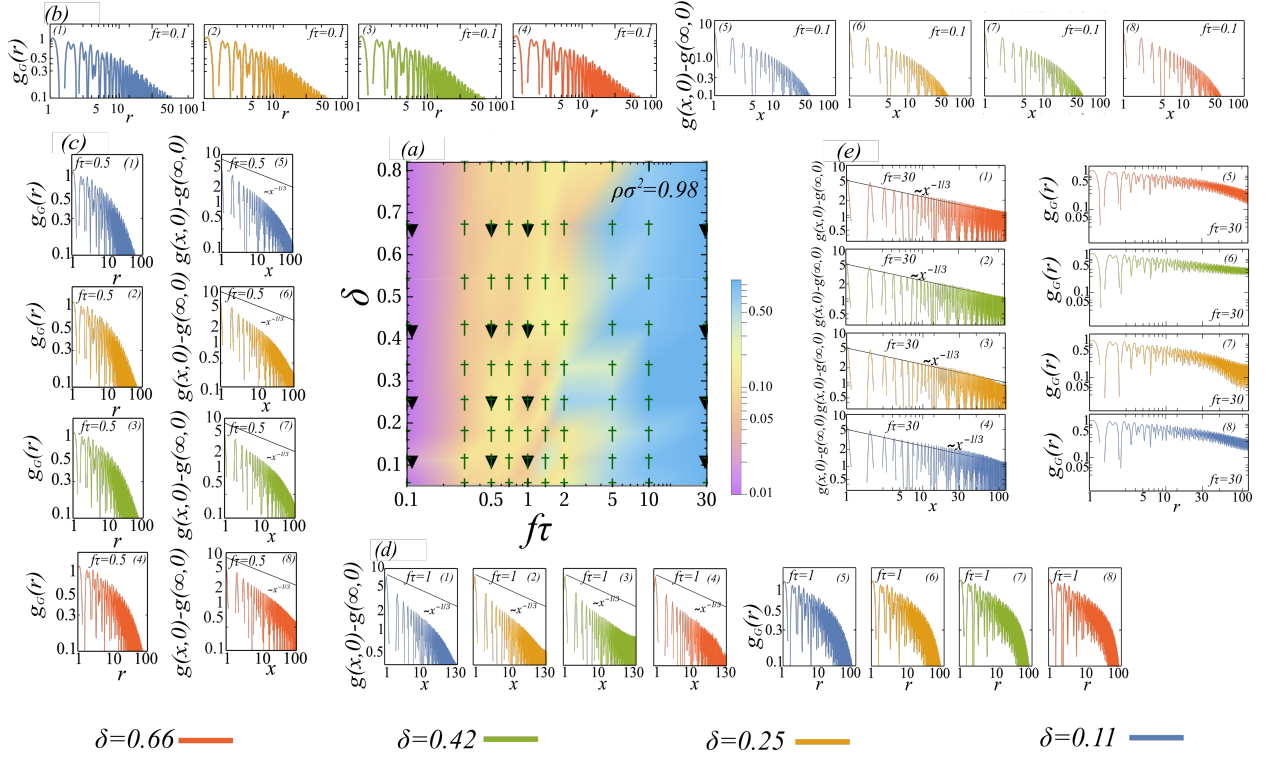}
\caption{Phase behavior in the asymmetry--frequency plane at
  $\rho\sigma^2=0.98$. Panel ({\bf a}) shows the time-averaged solid
  order parameter $\langle\psi_G\rangle$ (color scale); symbols
  indicate simulated state points. The surrounding panels display the
  positional correlation function $g_G(r)$ and the pair-correlation
  cut $g(x,0)-g(\infty,0)$ at selected frequencies: \textbf{(b)}
  $f\tau=0.1$, \textbf{(c)} $0.5$, \textbf{(d)} $1$, and \textbf{(e)}
  $30$. In each group, panels $1$--$4$ show $g_G(r)$ and panels
  $5$--$8$ show the corresponding pair-correlation cuts for
  $\delta=0.11$, $0.25$, $0.42$, and $0.67$, respectively. At low and
  intermediate frequencies ($f\tau\lesssim1$), $\langle\psi_G\rangle$
  remains small and positional correlations are short-ranged,
  consistent with density-modulated hexatic states [see also
  \cref{fig:psi6_vs_freq,fig:hexatic_density_plot}]. At high frequency
  ($f\tau=30$), $g_G(r)$ and $g(x,0)-g(\infty,0)$ exhibit power-law
  decay with $\eta_G\leq1/3$, accompanied by a large
  $\langle\psi_G\rangle$, indicating quasi-long-ranged positional
  order. The system thus crosses over from a density-modulated hexatic
  state to a solid-like state with increasing driving
  frequency.}
  \label{fig:phase1}
\end{figure*}
\begin{figure*}[!htbp]
\centering
\includegraphics[width=1\linewidth]{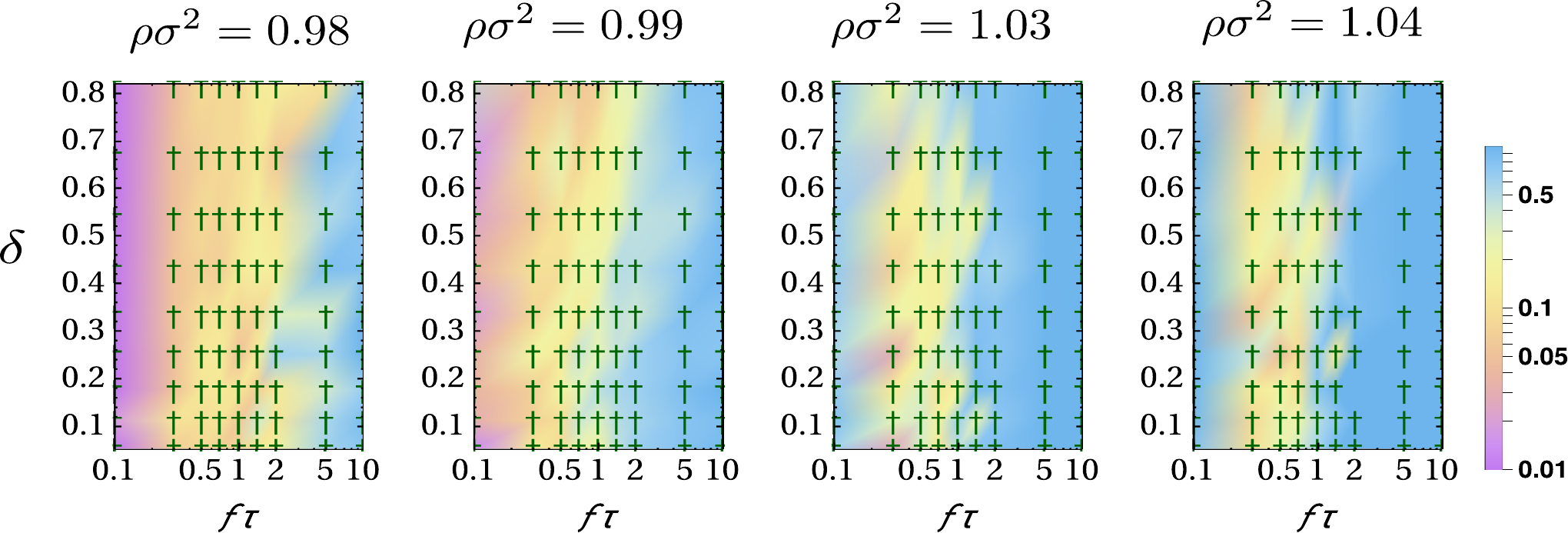}
\caption{Evolution of the nonequilibrium phase diagram with
  density. The color scale shows the solid order parameter
  $\langle\psi_{\bf G}\rangle$ in the frequency--asymmetry plane for
  $\rho\sigma^2=0.98$, $0.99$, $1.03$, and $1.04$; crosses denote
  simulated state points. Near the equilibrium hexatic melting
  density, a broad hexatic region occupies low and intermediate
  frequencies, while the solid phase emerges only at high
  frequency. With increasing density, the high-frequency solid region
  expands and the hexatic regime contracts. Above the equilibrium
  melting density, the system is solid at both low and high
  frequencies, with the hexatic phase restricted to an
  intermediate-frequency band.  }
  \label{fig:phase3}
\end{figure*}

The complete heat map of $\langle \psi_6 \rangle$ in the frequency -
asymmetry parameter plance for the density $\rho \sigma^2=0.98$ is
shown in \cref{fig:hexatic_density_plot}(a). $\langle\psi_6\rangle$
increases mainly along the frequency direction, while its variation
along the $\delta$ direction is comparatively weak. The cuts in
\cref{fig:hexatic_density_plot}(b) make this more explicit: for both
$\delta=0.11$ and $0.67$, the orientational order increases gradually
with frequency and the two curves remain close to each other. At the
higher density, $\rho\sigma^2=1.04$, shown in
\cref{fig:hexatic_density_plot}(c), the hexatic order is larger and
varies only weakly over the frequency range shown. Thus the asymmetry
parameter, although important for the current amplitude, does not by
itself select the hexatic order of the suspension.

How the positional order enters this pricture is qualitatively
illustrated in the time-averaged density profiles in
\cref{fig:rhoxy}. For $\rho\sigma^2=0.98$, the profiles at low and
intermediate frequencies are dominated by the density modulation
imposed by the external potential. The transverse density cuts show
only weak signatures of a triangular lattice in this regime. A
clearer, though still soft, triangular pattern appears only at a
higher frequency $f\tau=10$. At $\rho\sigma^2=1.04$, the situation
changes. A triangular structure is visible at low frequency and
becomes pronounced again at high frequency, whereas at $f\tau=0.5$ the
density pattern is much more strongly modulated along the drive
direction and the order is weakened in the transverse direction. These
profiles already indicate a re-entrant structural response at high
density: ordered at low frequency, melted at intermediate frequency,
and ordered again at high frequency.

The qualitative picture is quantified in the phase diagram shown in
\cref{fig:phase1}. The central panel gives the solid order parameter
$\langle\psi_{\bf G}\rangle$ in the $\delta$--$f\tau$ plane at
$\rho\sigma^2=0.98$, while the surrounding panels show the
corresponding pair-correlation cuts and positional correlation
functions at selected state points. Alongside the solid order
parameter we use the cut of the pair correlation function
$g(x,y)=\langle \rho(x,y)\rho(0,0)\rangle/\langle \rho^2 \rangle $ and
positional correlation function
$g_G(r)=\langle e^{\mathbf{G}.\mathbf{r}_{ij}}\delta(r-r_{ij})\rangle$
to identify the stable solid phase (see
\cref{appendix:equilibrium_melting} for more details). At low and
intermediate frequencies, the values of $\langle\psi_G\rangle$ are
small and $g_G(r)$ and $g(x,0)-g(\infty,0)$ decays rapidly (panels
(a), (b) and (c) in \cref{fig:phase1}), showing that positional order
is short-ranged.  However, the orientational order remains
appreciable, as shown above by $\langle\psi_6\rangle$ and by the
hexatic correlation function in
\cref{appendix:hexatic_corr_appendix}. The state is therefore a
density-modulated hexatic rather than an isotropic liquid. At high
frequency, the positional correlations become much longer ranged. In
particular, the cuts of the pair correlation function and the
positional correlation $g_G(r)$ show a power-law decay with an
exponent consistent with $\eta_G\leq\eta_G^*=1/3$, indicating
quasi-long-ranged positional order. Thus, at $\rho\sigma^2=0.98$, the
driven system evolves from a hexatic state at low and intermediate
frequencies to a solid-like state at sufficiently high frequency.

At the higher density, $\rho\sigma^2=1.04$, the corresponding phase
diagram and correlation functions are shown in
\cref{appendix:phase104}. Since this density is above the equilibrium
melting density, the system already has strong positional order at low
frequency. On increasing the driving frequency, the intermediate
frequency window shows short-ranged positional correlations while the
hexatic correlations remain long-ranged or slowly decaying. The system
therefore enters a hexatic state rather than an isotropic liquid. At
still higher frequencies, positional order is restored. The
high-density suspension therefore displays a re-entrant
solid--hexatic--solid sequence as the ratcheting frequency is
increased.

The density evolution of this behaviour is summarized in
\cref{fig:phase3}. At $\rho\sigma^2=0.98$, the hexatic region extends
over a broad range of low and intermediate frequencies, and the solid
phase appears only at high frequency. As the density is increased, the
high-frequency solid region expands, and the hexatic band narrows from
the high-frequency side. Above the equilibrium melting density, the
low-frequency solid also reappears, leaving the hexatic phase confined
to an intermediate-frequency window. Conversely, below the hexatic
melting regime, the low-frequency hexatic order is lost first, and at
still lower densities even the high-frequency drive can no longer
sustain appreciable orientational order. Furthermore, for densities
above the equilibrium solid melting point, with increasing asymmetry
parameter we notice pockets of parameter values with QLRO solid within
a parameter range that otherwise shows hexatic order, suggesting
multiple renentrant hexatic-solid transitions.

The phase behaviour therefore complements the observations on how the
two transport protocols act differently. Changing the driving
frequency moves the suspension between different structural regimes
and can create or destroy positional order.  Changing the asymmetry
parameter has a much weaker effect on the phase boundaries. It can
therefore be used to increase the directed current after the
structural state has been selected by density and frequency.

\FloatBarrier

\subsection{Defect Formation}
\label{sec:defect}
Finally, we examine defect formation across the structural phases and
transport regimes induced by the ratcheting drive. In crystalline
solids, the loss of positional order is accompanied by the
proliferation of topological defects. We identify defects using the
Voronoi coordination number $n_i$: particles with $n_i=6$ are locally
crystalline, whereas those with $n_i\neq 6$ are classified as
defects. Bound $5$--$7$ pairs correspond to dislocations, while
extended connected groups of non-sixfold coordinated particles form
defect clusters or grain-boundary-like structures. Representative
configurations are shown in \cref{appendix:defect_vis}.

At $\rho\sigma^2=0.98$, near the hexatic melting regime, defects are
abundant and occur primarily as bound dislocation pairs, quartets, and
extended chains of connected $5$--$7$ pairs, while free disclinations
remain rare. By contrast, at $\rho\sigma^2=1.04$, defect formation is
strongly suppressed: the extended chains largely disappear, and the
remaining defects consist predominantly of short chains and bound
quartets.

\begin{figure}[!ht]
\centering
\includegraphics[width=0.75\linewidth]{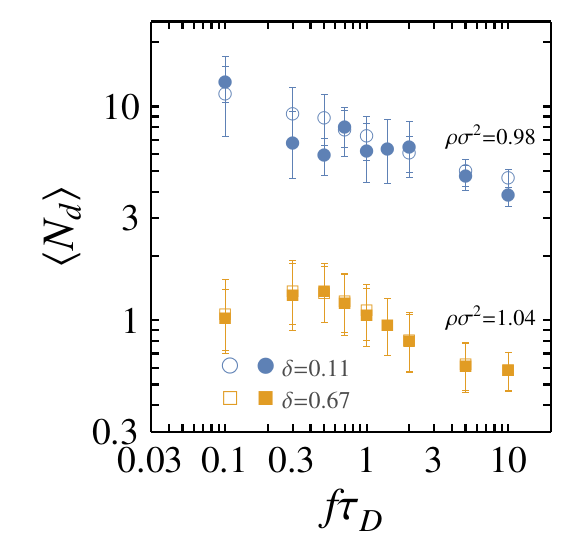}
\caption{Mean defect percentage $\langle N_d\rangle$ as a function of
  driving frequency for different densities and asymmetry parameters.
  At $\rho\sigma^2=0.98$, the defect fraction decreases with
  increasing frequency. At $\rho\sigma^2=1.04$, defects are suppressed
  except near the intermediate-frequency melting regime.}
\label{fig:defect_freq}
\end{figure}

The mean percentage of defects is
\begin{equation}
\langle N_d\rangle=\left(1-\frac{\langle n_6\rangle}{N}\right)\times 100,
\end{equation}
where $n_6$ is the number of particles with six Voronoi neighbours. As
shown in \cref{fig:defect_freq}, the defect fraction at
$\rho\sigma^2=0.98$ decreases as the driving frequency is increased,
consistent with the development of high-frequency positional order. At
$\rho\sigma^2=1.04$, the overall defect fraction is much smaller, but it
has a maximum in the intermediate-frequency window where the solid melts
into a hexatic state.

\begin{figure}[!ht]
\centering
\includegraphics[width=0.75\linewidth]{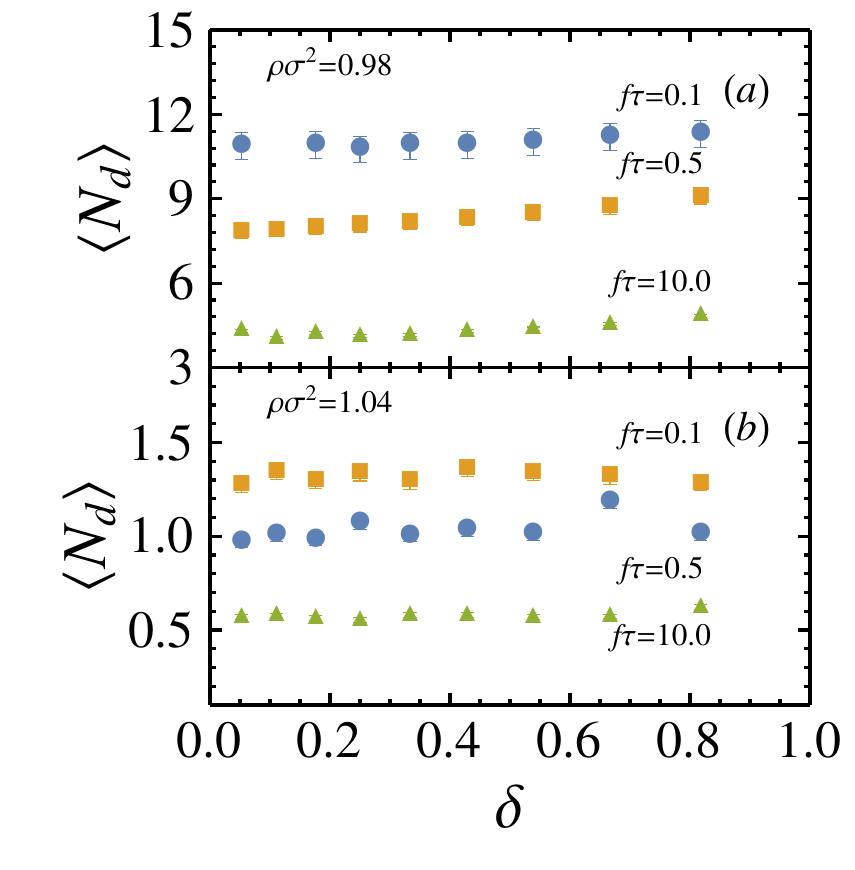}
\caption{Mean defect percentage $\langle N_d\rangle$ as a function of
  the asymmetry parameter for the densities and driving frequencies
  indicated in the figure. The defect fraction depends only weakly on
  $\delta$, showing that defect formation is controlled mainly by
  density and driving frequency.}
\label{fig:defect_beta}
\end{figure}

The weak dependence of the defect percentage on $\delta$
(\cref{fig:defect_beta}) provides a microscopic manifestation of the
order--current decoupling discussed above. Defect formation is
controlled primarily by density and driving frequency, whereas
asymmetry introduces only a directed bias. Consequently, the two
control parameters remain distinct at the level of defect dynamics:
frequency tuning couples transport to structural rearrangement, while
asymmetry tuning modulates the current with minimal impact on the
ordered or hexatic state.

\section{Outlook}
\label{sec_outlook}
We have investigated directed transport in a two-dimensional
suspension of repulsively interacting colloids driven by a stochastic
asymmetric flashing ratchet and identified two fundamentally distinct
mechanisms for controlling particle currents. When the ratchet
asymmetry is held fixed, the current exhibits a resonance as a
function of driving frequency, reflecting the collective relaxation
dynamics of the interacting suspension. By contrast, at fixed
frequency, increasing the asymmetry of the ratchet primarily
strengthens the directed bias and can substantially enhance the
current. Although both parameters increase transport, they do so
through different physical pathways.

This distinction becomes evident in the accompanying structural
response. Frequency tuning couples directly to the internal dynamics
of the suspension and drives pronounced structural
reorganization. Depending on density, the system displays
density-modulated hexatic states and, above the equilibrium melting
density, a re-entrant solid–hexatic–solid sequence as the driving
frequency is increased. These transitions are mediated by the creation
and annihilation of topological defects: the hexatic regime is
characterized by short-ranged positional order coexisting with robust
sixfold orientational order and an enhanced population of dislocations
and defect clusters, whereas higher frequencies suppress defect
formation and restore crystalline order. In contrast, varying the
ratchet asymmetry over the same range produces comparatively little
change in the structural state, indicating that asymmetry acts
predominantly as a transport-control parameter rather than a
structural one.

Taken together, our results show that transport and organization in
driven colloidal suspensions can be controlled independently through
distinct external parameters. Frequency tuning and asymmetry tuning
provide complementary means of manipulating nonequilibrium steady
states: the former couples transport to defect-mediated structural
evolution, while the latter largely regulates transport without
significantly altering orientational order. This separation of roles
is particularly striking near the equilibrium hexatic-melting regime,
where substantial directed currents can be generated while preserving
a high degree of sixfold order. More broadly, these findings highlight
how time-dependent driving protocols can be used to independently
engineer transport and structure in interacting many-body systems,
opening new possibilities for the design of driven soft materials and
colloidal transport devices.

Our predictions can be tested in colloidal systems where
time-dependent asymmetric potentials are realized using optical,
electrokinetic, or magnetic control~\cite{Wei1998, Tierno2012,
  Faucheux1995, Cubero2016, Thomas2009}. Holographic optical tweezers
and spatial light modulator–based setups already make it possible to
program flashing ratchets with independently tunable spatial asymmetry
and driving protocols~\cite{tang2021}, and existing experiments on
optically driven Brownian ratchets have demonstrated controlled
directed transport using programmable switching and feedback.

A particularly clean test would be to use dense colloidal monolayers
near the hexatic regime, where one can independently vary the drive
frequency and ratchet asymmetry while tracking both transport and
hexatic order in real time. This would allow a direct check of the
predicted separation between enhanced current and structural ordering.

More generally, these platforms offer a way to design nonequilibrium
states where transport can be boosted without destroying crystalline
order, pointing toward programmable colloidal materials with
independently controllable flow and structure.

\section*{Acknowledgments}
Debasish Chaudhuri acknowledges financial support from the Department of Atomic
Energy (DAE) through Grant No. 1603/2/2020/IoP/R\&D-II/15028, a
Visiting Professorship at CY Cergy Paris Universit{\'e}, and an
Associateship of IIT Bombay.
Dipanjan Chakraborty acknowledges financial support from DST-SERB through
grant No. CRG/2021/003640.
\bibliographystyle{unsrt}
\bibliography{phase_linear_asymmetric_ratchet_cited.bib}

@Book{		  cubero2016,
  author	= {D. Cubero and F. Renzoni},
  title		= {Brownian Ratchets: From Statistical Physics to Bio and
		  Nano-Motors},
  publisher	= {Cambridge University Press},
  year		= {2016}
}

@Article{	  khali2020structure,
  title		= {A structure--dynamics relationship in ratcheted colloids:
		  resonance melting, dislocations, and defect clusters},
  author	= {Khali, Shubhendu Shekhar and Chakraborty, Dipanjan and
		  Chaudhuri, Debasish},
  journal	= {Soft Matter},
  volume	= {16},
  number	= {10},
  pages		= {2552--2564},
  year		= {2020},
  publisher	= {Royal Society of Chemistry}
}

@Misc{		  frenkel2002algorithms,
  title		= {From algorithms to applications},
  author	= {Frenkel, Daan and Smit, Berend},
  year		= {2002},
  publisher	= {New York, NY: Academic Press}
}

@Article{	  rossini2018sliding,
  title		= {Sliding states of a soft-colloid cluster crystal: Cluster
		  versus single-particle hopping},
  author	= {Rossini, Mirko and Consonni, Lorenzo and Stenco, Andrea
		  and Reatto, Luciano and Manini, Nicola},
  journal	= {Physical Review E},
  volume	= {97},
  number	= {5},
  pages		= {052614},
  year		= {2018},
  publisher	= {APS}
}

@Article{	  martin2025colloidal,
  title		= {Colloidal Model for Investigating Optimal Efficiency in
		  Weakly Coupled Ratchet Motors},
  author	= {Mart{\'\i}n-Roca, Jos{\'e} and Izquierdo Solis, Laura and
		  Mart{\'\i}nez Pedrero, Fernando and Casadejust, Pau and
		  Pagonabarraga, Ignacio and Calero, Carles},
  journal	= {Physical Review Letters},
  volume	= {135},
  number	= {2},
  pages		= {028301},
  year		= {2025},
  publisher	= {APS}
}

@Article{	  patil2022using,
  title		= {Using the thermal ratchet mechanism to achieve net
		  motility in magnetic microswimmers},
  author	= {Patil, Gouri and Mandal, Pranay and Ghosh, Ambarish},
  journal	= {Physical Review Letters},
  volume	= {129},
  number	= {19},
  pages		= {198002},
  year		= {2022},
  publisher	= {APS}
}

@Unpublished{	  mandal2025u,
  author	= {Mandal, Sudipta and Chakraborty, Dipanjan and Chaudhuri,
		  Debasish },
  title		= {Impact of Asymmetry on Resonance Frequency and Current in
		  a Stochastic Flashing Ratchet},
  note		= {Manuscript in preparation},
  year		= {2025},
  month		= {December}
}

@Article{	  tang2021,
  title		= {Experimental demonstration of optical Brownian ratchet by
		  controllable phase profile of light},
  journal	= {Optics and Lasers in Engineering},
  volume	= {145},
  pages		= {106671},
  year		= {2021},
  issn		= {0143-8166},
  doi		= {https://doi.org/10.1016/j.optlaseng.2021.106671},
  author	= {Xionggui Tang and Yi Shen and Yanhua Xu}
}

@Article{	  thomas2009,
  title		= {Experimental investigation of a Brownian ratchet effect in
		  ferrofluids},
  author	= {John, Thomas and Stannarius, Ralf},
  journal	= {Phys. Rev. E},
  volume	= {80},
  issue		= {5},
  pages		= {050104(R)},
  numpages	= {4},
  year		= {2009},
  month		= {Nov},
  publisher	= {American Physical Society},
  doi		= {10.1103/PhysRevE.80.050104},
  url		= {https://link.aps.org/doi/10.1103/PhysRevE.80.050104}
}

@Article{	  savelev2003,
  author	= {Savel'ev, Sergey and Marchesoni, F. and Nori, Franco},
  doi		= {10.1103/PhysRevLett.91.010601},
  issn		= {0031-9007},
  journal	= {Phys. Rev. Lett.},
  month		= {jul},
  number	= {1},
  pages		= {010601},
  title		= {{Controlling Transport in Mixtures of Interacting
		  Particles using Brownian Motors}},
  url		= {https://link.aps.org/doi/10.1103/PhysRevLett.91.010601},
  volume	= {91},
  year		= {2003}
}

@Article{	  savelev2004,
  author	= {Savel'ev, Sergey and Marchesoni, Fabio and Nori, Franco},
  doi		= {10.1103/PhysRevE.70.061107},
  issn		= {1539-3755},
  journal	= {Phys. Rev. E},
  month		= {dec},
  number	= {6},
  pages		= {061107},
  title		= {{Stochastic transport of interacting particles in
		  periodically driven ratchets}},
  url		= {http://link.aps.org/doi/10.1103/PhysRevE.70.061107
		  https://link.aps.org/doi/10.1103/PhysRevE.70.061107},
  volume	= {70},
  year		= {2004}
}

@Article{	  pototsky2010,
  author	= {Pototsky, A. and Archer, A. J. and Bestehorn, M. and
		  Merkt, D. and Savel'ev, S. and Marchesoni, F.},
  doi		= {10.1103/PhysRevE.82.030401},
  issn		= {1539-3755},
  journal	= {Phys. Rev. E},
  month		= {sep},
  number	= {3},
  pages		= {030401},
  title		= {{Collective shuttling of attracting particles in
		  asymmetric narrow channels}},
  url		= {https://link.aps.org/doi/10.1103/PhysRevE.82.030401},
  volume	= {82},
  year		= {2010}
}

@Article{	  chaudhuri2015f,
  author	= {Chaudhuri, Debasish},
  doi		= {10.1088/1742-6596/638/1/012011},
  issn		= {1742-6588},
  journal	= {J. Phys. Conf. Ser.},
  month		= {sep},
  number	= {1},
  pages		= {012011},
  title		= {{Stochastic models of classical particle pumps : Density
		  dependence of directed current}},
  url		= {http://stacks.iop.org/1742-6596/638/i=1/a=012011?key=crossref.985b1c7eca818a5e06df8182364ccade},
  volume	= {638},
  year		= {2015}
}

@Article{	  chaudhuri2015,
  archiveprefix	= {arXiv},
  arxivid	= {1409.6966},
  author	= {Chaudhuri, Debasish and Raju, Archishman and Dhar,
		  Abhishek},
  doi		= {10.1103/PhysRevE.91.050103},
  eprint	= {1409.6966},
  issn		= {1539-3755},
  journal	= {Phys. Rev. E},
  month		= {may},
  number	= {5},
  pages		= {050103},
  pmid		= {26066100},
  title		= {{Pumping single-file colloids: Absence of current
		  reversal}},
  url		= {https://link.aps.org/doi/10.1103/PhysRevE.91.050103},
  volume	= {91},
  year		= {2015}
}

@Article{	  chakraborty2014,
  author	= {Chakraborty, Dipanjan and Chaudhuri, Debasish},
  doi		= {10.1103/PhysRevE.91.050301},
  journal	= {Physical Review E - Statistical, Nonlinear, and Soft
		  Matter Physics},
  pages		= {050301(R)},
  title		= {{Stochastic ratcheting of two-dimensional colloids :
		  Directed current and dynamical transitions}},
  url		= {http://journals.aps.org/pre/pdf/10.1103/PhysRevE.91.050301},
  volume	= {91},
  year		= {2015}
}

@Article{	  astumian2002,
  author	= {Astumian, R. D. and H\"{a}nggi, Peter},
  title		= {{Brownian Motors}},
  journal	= {Physics Today},
  year		= {2002},
  volume	= {55},
  pages		= {33},
  number	= {11},
  publisher	= {Elsevier}
}

@Article{	  gadsby2009,
  author	= {Gadsby, David C. and Takeuchi, Ayako and Artigas, Pablo
		  and Reyes, Nicol{\'{a}}s},
  doi		= {10.1098/rstb.2008.0243},
  issn		= {0962-8436},
  journal	= {Philos. Trans. R. Soc. B Biol. Sci.},
  month		= {jan},
  number	= {1514},
  pages		= {229--238},
  title		= {{Peering into an ATPase ion pump with single-channel
		  recordings}},
  url		= {https://royalsocietypublishing.org/doi/10.1098/rstb.2008.0243},
  volume	= {364},
  year		= {2009}
}

@Article{	  chaudhuri2011,
  author	= {Chaudhuri, Debasish and Dhar, Abhishek},
  title		= {{Stochastic pump of interacting particles}},
  journal	= {EPL (Europhysics Letters)},
  year		= {2011},
  volume	= {94},
  pages		= {30006},
  number	= {3},
  month		= may,
  doi		= {10.1209/0295-5075/94/30006},
  issn		= {0295-5075},
  url		= {http://stacks.iop.org/0295-5075/94/i=3/a=30006?key=crossref.acd7414e0bf43adc4a715580a246b136}
}

@Article{	  chaudhuri2006,
  author	= {Chaudhuri, Debasish and Sengupta, Surajit},
  title		= {{Direct test of defect-mediated laser-induced melting
		  theory for two-dimensional solids}},
  journal	= {Physical Review E},
  year		= {2006},
  volume	= {73},
  pages		= {11507},
  number	= {1},
  doi		= {10.1103/PhysRevE.73.011507},
  publisher	= {APS},
  url		= {http://link.aps.org/doi/10.1103/PhysRevE.73.011507}
}

@Article{	  marathe2008,
  archiveprefix	= {arXiv},
  arxivid	= {0809.2468},
  author	= {Marathe, Rahul and Jain, Kavita and Dhar, Abhishek},
  doi		= {10.1088/1742-5468/2008/11/P11014},
  eprint	= {0809.2468},
  journal	= {J. Stat. Mech. Theory Exp.},
  month		= {sep},
  pages		= {P11014},
  publisher	= {Institute of Physics Publishing},
  title		= {{Particle current in symmetric exclusion process with
		  time-dependent hopping rates}},
  url		= {http://www.iop.org/EJ/abstract/1742-5468/2008/11/P11014},
  volume	= {2008},
  year		= {2008}
}

@Article{	  jain2007,
  archiveprefix	= {arXiv},
  arxivid	= {cond-mat/0703335},
  author	= {Jain, Kavita and Marathe, Rahul and Chaudhuri, Abhishek
		  and Dhar, Abhishek},
  doi		= {10.1103/PhysRevLett.99.190601},
  eprint	= {0703335},
  issn		= {0031-9007},
  journal	= {Phys. Rev. Lett.},
  month		= {nov},
  number	= {19},
  pages		= {190601},
  primaryclass	= {cond-mat},
  title		= {{Driving Particle Current through Narrow Channels Using a
		  Classical Pump}},
  url		= {http://link.aps.org/doi/10.1103/PhysRevLett.99.190601},
  volume	= {99},
  year		= {2007}
}

@Article{	  citro2003,
  author	= {Citro, R. and Andrei, N. and Niu, Q.},
  doi		= {10.1103/PhysRevB.68.165312},
  issn		= {0163-1829},
  journal	= {Phys. Rev. B},
  month		= {oct},
  number	= {16},
  pages		= {165312},
  title		= {{Pumping in an interacting quantum wire}},
  url		= {http://link.aps.org/doi/10.1103/PhysRevB.68.165312},
  volume	= {68},
  year		= {2003}
}

@Article{	  derenyi1996,
  author	= {Der\'{e}nyi, Imre and Ajdari, Armand},
  title		= {{Collective transport of particles in a "flashing"
		  periodic potential}},
  journal	= {Physical Review E},
  year		= {1996},
  volume	= {54},
  pages		= {R5--R8},
  number	= {1},
  month		= jul,
  doi		= {10.1103/PhysRevE.54.R5},
  issn		= {1063-651X},
  publisher	= {American Physical Society},
  shorttitle	= {Phys. Rev. E},
  url		= {http://link.aps.org/doi/10.1103/PhysRevE.54.R5}
}

@Article{	  derenyi1995,
  author	= {Der\'{e}nyi, Imre and Vicsek, T},
  title		= {{Cooperative transport of Brownian particles}},
  journal	= {Physical review letters},
  year		= {1995},
  volume	= {75},
  pages		= {374},
  url		= {http://link.aps.org/doi/10.1103/PhysRevLett.75.374}
}

@Article{	  faucheux1995,
  author	= {Faucheux, L. and Bourdieu, L. and Kaplan, P. and
		  Libchaber, A.},
  title		= {{Optical Thermal Ratchet}},
  journal	= {Physical Review Letters},
  year		= {1995},
  volume	= {74},
  pages		= {1504--1507},
  number	= {9},
  month		= feb,
  doi		= {10.1103/PhysRevLett.74.1504},
  issn		= {0031-9007},
  url		= {http://link.aps.org/doi/10.1103/PhysRevLett.74.1504}
}

@Article{	  hanggi2009,
  author	= {H\"{a}nggi, Peter},
  title		= {{Artificial Brownian motors: Controlling transport on the
		  nanoscale}},
  journal	= {Reviews of Modern Physics},
  year		= {2009},
  volume	= {81},
  pages		= {387--442},
  number	= {1},
  month		= mar,
  doi		= {10.1103/RevModPhys.81.387},
  issn		= {0034-6861},
  url		= {http://link.aps.org/doi/10.1103/RevModPhys.81.387}
}

@Article{	  julicher1997a,
  author	= {Julicher, Frank and Ajdari, Armand and Prost, Jacques},
  title		= {{Modeling molecular motors}},
  journal	= {Reviews of Modern Physics},
  year		= {1997},
  volume	= {69},
  pages		= {1269--1282},
  number	= {4},
  month		= oct,
  doi		= {10.1103/RevModPhys.69.1269},
  issn		= {0034-6861},
  publisher	= {American Physical Society},
  shorttitle	= {Rev. Mod. Phys.},
  url		= {http://link.aps.org/doi/10.1103/RevModPhys.69.1269}
}

@Article{	  leibler1994,
  author	= {S. Leibler},
  title		= {Moving forward noisily},
  journal	= {Nature},
  year		= {1994},
  volume	= {370},
  pages		= {412},
  quality	= {1}
}

@Article{	  lopez2008,
  author	= {Lopez, Benjamin and Kuwada, Nathan and Craig, Erin and
		  Long, Brian and Linke, Heiner},
  title		= {{Realization of a Feedback Controlled Flashing Ratchet}},
  journal	= {Physical Review Letters},
  year		= {2008},
  volume	= {101},
  pages		= {220601},
  number	= {22},
  month		= nov,
  doi		= {10.1103/PhysRevLett.101.220601},
  issn		= {0031-9007},
  url		= {http://link.aps.org/doi/10.1103/PhysRevLett.101.220601}
}

@Article{	  marquet2002,
  author	= {Marquet, C. and Buguin, A. and Talini, L. and Silberzan,
		  P.},
  title		= {{Rectified Motion of Colloids in Asymmetrically Structured
		  Channels}},
  journal	= {Physical Review Letters},
  year		= {2002},
  volume	= {88},
  pages		= {168301},
  number	= {16},
  month		= apr,
  doi		= {10.1103/PhysRevLett.88.168301},
  issn		= {0031-9007},
  url		= {http://link.aps.org/doi/10.1103/PhysRevLett.88.168301}
}

@Article{	  reimann2002,
  author	= {Reimann, P},
  title		= {{Brownian motors: noisy transport far from equilibrium}},
  journal	= {Physics Reports},
  year		= {2002},
  volume	= {361},
  pages		= {57--265},
  number	= {2-4},
  month		= apr,
  doi		= {10.1016/S0370-1573(01)00081-3},
  issn		= {03701573},
  url		= {http://dx.doi.org/10.1016/S0370-1573(01)00081-3}
}

@Article{	  rousselet1994,
  author	= {Rousselet, J and Salome, L and Ajdari, A and Prost, J},
  title		= {{Directional motion of Brownian particles induced by a
		  periodic asymmetric potential}},
  journal	= {Nature},
  year		= {1994},
  volume	= {370},
  pages		= {446},
  url		= {http://phstudy.technion.ac.il/~sp116029/references/Project
		  No.18/Rousselet.pdf}
}

@Article{	  tierno2012,
  author	= {Tierno, Pietro},
  title		= {{Depinning and Collective Dynamics of Magnetically Driven
		  Colloidal Monolayers}},
  journal	= {Physical Review Letters},
  year		= {2012},
  volume	= {109},
  pages		= {198304},
  number	= {19},
  month		= nov,
  doi		= {10.1103/PhysRevLett.109.198304},
  issn		= {0031-9007},
  url		= {http://link.aps.org/doi/10.1103/PhysRevLett.109.198304}
}

@Article{	  tierno2010,
  author	= {Tierno, Pietro and Reimann, Peter and Johansen, Tom H. and
		  Sagu\'{e}s, Francesc},
  title		= {{Giant Transversal Particle Diffusion in a Longitudinal
		  Magnetic Ratchet}},
  journal	= {Physical Review Letters},
  year		= {2010},
  volume	= {105},
  pages		= {230602},
  number	= {23},
  month		= dec,
  doi		= {10.1103/PhysRevLett.105.230602},
  issn		= {0031-9007},
  url		= {http://link.aps.org/doi/10.1103/PhysRevLett.105.230602}
}

@Article{	  wei1998,
  author	= {Wei, Q.-H. and Bechinger, Clemens and Rudhardt, D. and
		  Leiderer, P.},
  title		= {{Experimental Study of Laser-Induced Melting in
		  Two-Dimensional Colloids}},
  journal	= {Physical Review Letters},
  year		= {1998},
  volume	= {81},
  pages		= {2606--2609},
  number	= {12},
  month		= sep,
  doi		= {10.1103/PhysRevLett.81.2606},
  issn		= {0031-9007},
  url		= {http://link.aps.org/doi/10.1103/PhysRevLett.81.2606}
}

@Article{	  kapfer:2015ca,
  author	= {Kapfer, Sebastian C and Krauth, Werner},
  title		= {{Two-Dimensional Melting: From Liquid-Hexatic Coexistence
		  to Continuous Transitions}},
  journal	= {Physical Review Letters},
  year		= {2015},
  volume	= {114},
  number	= {3},
  pages		= {035702--5},
  month		= jan
}

@article{Diwakar2024,
  author  = {Diwakar, Nidhi M. and Yossifon, Gilad and Miloh, Touvia and Velev, Orlin D.},
  title   = {Active microparticle propulsion pervasively powered by asymmetric {AC} field electrophoresis},
  journal = {Journal of Colloid and Interface Science},
  volume  = {676},
  pages   = {817--825},
  year    = {2024},
  doi     = {10.1016/j.jcis.2024.07.141}
}

@article{Herman2023,
  title = {Ratchet-Based Ion Pumps for Selective Ion Separations},
  author = {Herman, Alon and Ager, Joel W. and Ardo, Shane and Segev, Gideon},
  journal = {PRX Energy},
  volume = {2},
  issue = {2},
  pages = {023001},
  numpages = {18},
  year = {2023},
  month = {Apr},
  publisher = {American Physical Society},
  doi = {10.1103/PRXEnergy.2.023001},
  url = {https://link.aps.org/doi/10.1103/PRXEnergy.2.023001}
}

@article{Molcrette2022,
author = {Bastien Molcrette  and Léa Chazot-Franguiadakis  and François Liénard  and Zsombor Balassy  and Céline Freton  and Christophe Grangeasse  and Fabien Montel },
title = {Experimental study of a nanoscale translocation ratchet},
journal = {Proceedings of the National Academy of Sciences},
volume = {119},
number = {30},
pages = {e2202527119},
year = {2022},
doi = {10.1073/pnas.2202527119},
URL = {https://www.pnas.org/doi/abs/10.1073/pnas.2202527119},
eprint = {https://www.pnas.org/doi/pdf/10.1073/pnas.2202527119},
}

@article{Wen2025,
    author = {Wen, Yan and Li, Zhihao and Wang, Haiqin and Zheng, Jing and Tang, Jinyao and Xu, Xinpeng and Lai, Pik-Yin and Tong, Penger},
    title = {Ratchet effect of self-propelled colloids in an asymmetric periodic potential},
    journal = {The Journal of Chemical Physics},
    volume = {162},
    number = {20},
    pages = {204903},
    year = {2025},
    month = {05},
    issn = {0021-9606},
    doi = {10.1063/5.0237153},
    url = {https://doi.org/10.1063/5.0237153},
    eprint = {https://pubs.aip.org/aip/jcp/article-pdf/doi/10.1063/5.0237153/20530181/204903_1_5.0237153.pdf},
}

\clearpage
\onecolumn
\begin{center}
 {\bf {\Large{Appendix}}}
 \end{center}

 \begin{appendices}
\counterwithin{figure}{section}

\section{Defect Visualisation}
\label{appendix:defect_vis}
\begin{figure*}[!ht]
  \centering
  \includegraphics[width=\textwidth]{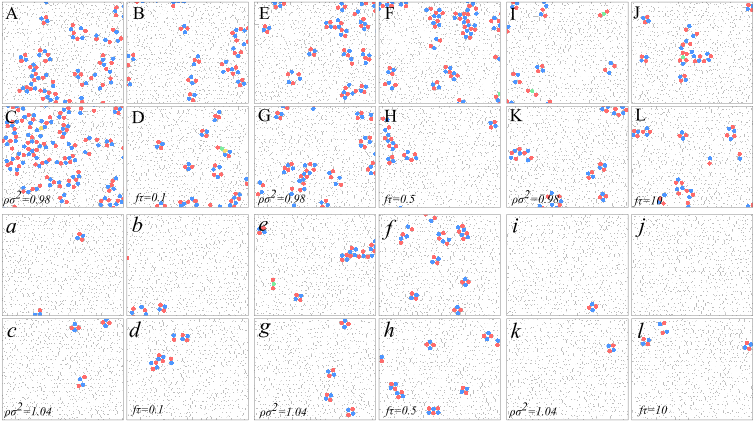}
  \caption{ Snapshots of representative subdomains highlighting the
    defect structures are shown for densities $\rho\sigma^{2}=0.98$
    (Figures~\textbf{A}--\textbf{L}) and $\rho\sigma^{2}=1.04$
    (Figures~\textbf{a}--\textbf{l}) across three driving frequencies:
    $f\tau = 0.1$ (Figures~\textbf{A}--\textbf{D},
    \textbf{a}--\textbf{d}), $0.5$ (Figures~\textbf{E}--\textbf{H},
    \textbf{e}--\textbf{h}), and $10$ (Figures~\textbf{I}--\textbf{L},
    \textbf{i}--\textbf{l}). At densities below the equilibrium
    melting point, the dominant defects are chains of dislocations
    (bound $5-7$ disclination pairs) together with defect quartets
    (bound $5--7--7--5$ dislocation pairs) and disclinations (isolated
    $5-$ and $7-$ fold defects) are few in number. As the
    frequency of the drive is increased the formation of defects is
    suppressed in the system. In contrast at densities above the
    equilibrium melting point, the formation of defects is drastically
    reduced (figures \textbf{a} -- \textbf{l}). While defect chains
    are observed in the system, they are shorter in length and few in
    numbers.}
  \label{fig:defect_visualisation}
\end{figure*}
\clearpage
\section{Phase diagram at density $\rho \sigma^2=1.04$}
\label{appendix:phase104}
In this section we show the dynamical phase diagram in the
frequency--asymmetry plane at a density of $\rho \sigma^2=1.04$. The
density is above the equilibrium melting point.
\begin{figure*}[!h]
\centering
\includegraphics[width=\linewidth]{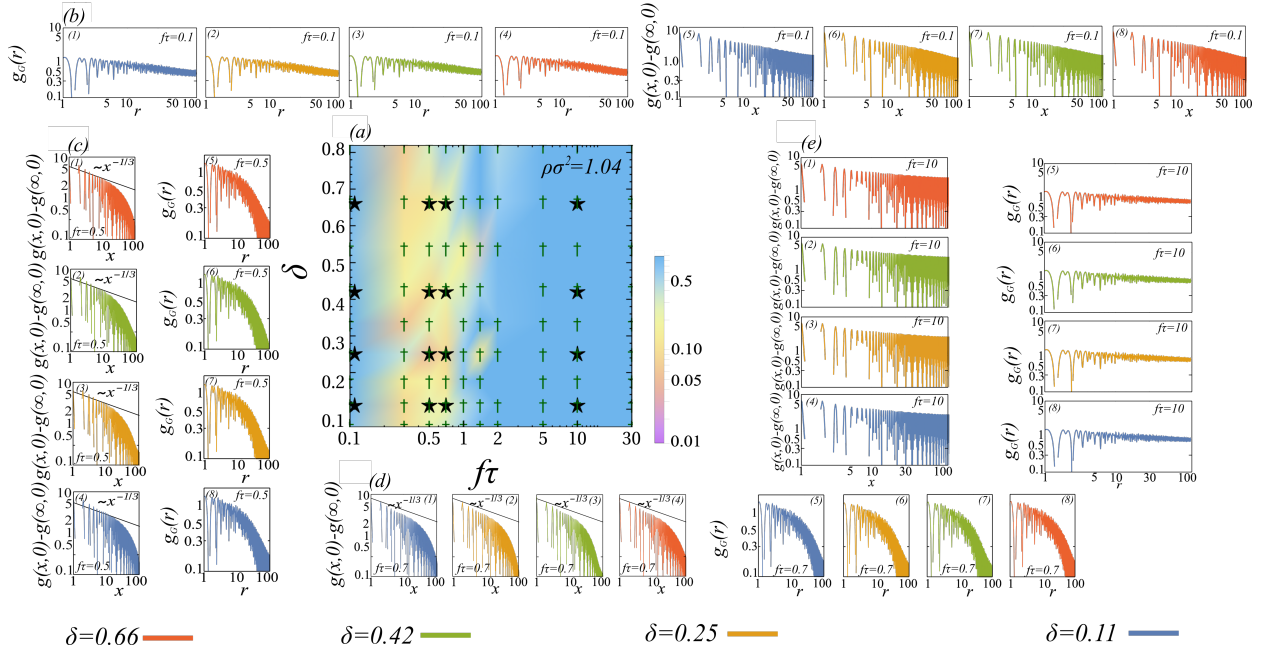}
\caption{Phase diagram in the asymmetric parameter($\delta$)-
  frequency($f$) plane for a fixed particle density
  $\rho\sigma^2=1.04$ is shown in \textbf{figure (a)}. The color code
  indicates the values of the average solid order parameter
  $\langle \psi_G\rangle$. In the rest of the figures we show the cut
  of the pair correlation function $g(x,0)-g(x,\infty)$, the
  positional correlation function $g_G(r)$ for three frequencies of
  the drive: low frequency $f \tau =0.1$ (figures (b-1)-- (b--8)), two
  intermediate frequencies $f \tau=0.5$ (figures (c-1)-- (c--8)) and
  $0.7$ (figures (d-1)-- (d--8) and one high frequency $f \tau=30$
  (figures (e-1)--(e-8)). For each of these frequencies (figure panels
  a,b,c,d) we show the result for four different values of $\delta$ as
  indicated: $\delta=0.11$, the numerals labeled $1$ and $5$ ,
  $\delta=0.25$, the numerals labeled $2$ and $6$, $\delta=0.42$, the
  numerals labeled $3$ and $7$ and $\delta=0.67$, the numerals $4$ and
  $8$.The wide band of hexatic phase observed in \cref{fig:phase1}(a)
  becomes significantly narrower, confined within the frequency range
  $0.2 \le f\tau \le 2$. While pockets of QLRO triangular solid is
  observed within this frequency range, beyond $f \tau \ge 2$ a stable
  solid is observed.  The cut of the pair correlation function and the
  positional correlation exhibits a power law decay with an exponent
  $\eta_G\le 1/3$ at a frequency of $f \tau =0.1$ (figures
  (b-1)--(b-8)) and at a frequency of $f \tau=10$ (figures
  (e-1)--(e-8)). In the intermediate frequency range $f \tau=0.5$ and
  $f\tau=0.7$ both $g_G(r)$ and $g(x,0)-g(\infty,0)$ shows short range
  decay indicating hexatic phase.The stars in the phase diagram
  (figure (e)) are the points for which data for $g(x,0)-g(\infty,0)$
  and $g_G(r)$ are shown in the surrounding panels.}
\label{fig:phase2}
\end{figure*}
% \twocolumngrid
\clearpage
\section{Hexatic correlation function}
\label{appendix:hexatic_corr_appendix}
\begin{figure}[!ht]
  \includegraphics[width=\linewidth]{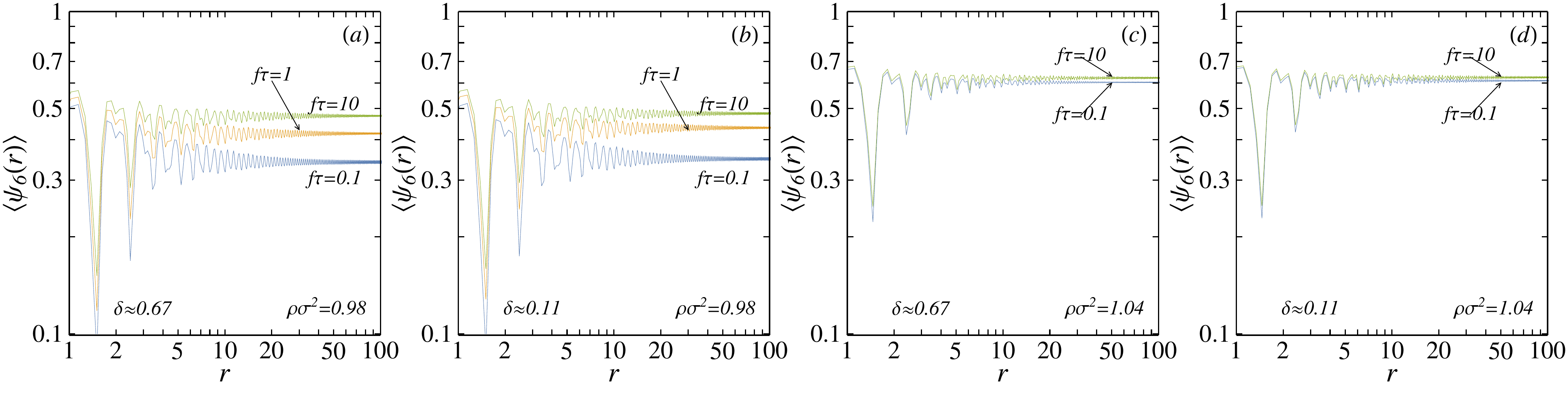}
  \label{fig:hexatic_corr_function}
  \caption{Plot of the hexatic correlation function for two densities
    $\rho \sigma^2=0.98$ (figures (a) and (b)) and $1.04$ (figures (c)
    and (d)) at three frequencies indicated in the figure.}
\end{figure}

The hexatic correlation function is constructed from the local hexatic
order parameter defined in \cref{eq:local_psi6}. The hexatic
correlation function is defined as
$g_6(r)=\langle {\psi_6^{i}}^* \psi_6^{j}\delta(r-r_{ij})\rangle$. In
\cref{fig:hexatic_corr_function}, we show the hexatic correlation function for
two densities $\rho \sigma^2=0.98$ (figures (a) and (b)) and
$\rho \sigma^2=1.04$ (figures (c) and (d)). For each density, we show
the spatial correlation for two values of asymmetry parameter as
indicated in the legend. From the correlation function it is clear the
orientational order in the system is long ranged and does not decay
due to the presence of the external potential.
\section{Equilibrium melting of the two-dimensional solid}
\label{appendix:equilibrium_melting}

A
two-dimensional solid undergoes a melting transition to an isotropic
liquid phase via an intermediate hexatic phase. The stable solid phase
has a quasi long range translational order and is characterized by an
algebraic decay of the positional correlation. In contrast the hexatic
correlation remains long ranged in the solid phase. In a stable
hexatic phase, the hexatic correlation function shows an algebraic
decay whereas the positional correlation shows a short range decay. In
the liquid phase, both the correlations exhibit a short range decay.
% We briefly outline the different physical quantities and correlation
% functions that has been used in the study and is later used to
% distinguish the different phases in the system.
We briefly summarize the various physical quantities and correlation
functions used in this study that is later employed to distinguish
qualitatively and quantitatively between the different phases of the
system.

\section*{Identifying the solid phase}
\subsubsection{Positional correlation}

Positional correlation is defined by
$g_{\vec{G}}(r)=\langle e^{\complexi \vec{G}\cdot \vec{r}_{ij}}\delta(r-r_{ij})\rangle$, where
$\vec{r}_{ij}=\vec{r}_i-\vec{r}_j$ is the inter-particle separation
vector, $r_{ij} = \mid r_{ij} \mid$, and $\vec{G}$ denotes the the reciprocal
lattice vectors corresponding to the six quasi- Bragg peaks at
$\vec{G}_1 = (0,\pm 2\pi/a_y)$ and and $\vec{G}_2 =(\pm 2\pi/a, \pm\pi/a_y)$. The
change in the decay in correlation from power law
$g_{\vec{G}}(r)\sim r^{-\eta_G}$ in the quasi-long-range order (QLRO) solid
phase. According to the KTHNY theory, this value corresponds to
$-\eta_G=1/3$. In the hexatic and liquid phases, the positional
correlation exhibits a short range decay.

\subsubsection{Cut of the pair correlation function}
The pair correlation function
$g(x,y)=\langle \rho(x,y)\rho(0,0)\rangle/\langle \rho\rangle^2$ is
yet another tool in identifying a solid, hexatic and a liquid
phase. In the solid phase, a distinct triangular lattice is
observed. As the solid melts, in the hexatic phase, the triangular
lattice structure gives way to diffused concentric hexagonal rings. In
the liquid phase, the hexagonal rings are replaced by concentric
circular rings characteristic of isotropic liquid phase.

While $g(x,y)$ provides a qualitative estimate of the
transition points, the more accurate estimate is obtained from the cut
of the pair correlation function along the $x$-direction. 
% The cut of the pair correlation function  along the $x$-direction
% provides a quantitative estimation of the melting transition.
In a stable solid phase, the positional order is quasi-long ranged and the
correlation $g(x,0)-g(\infty,0)$ decays algebraically with an exponent
$\eta_G \le 1/3$, with $\eta^{*}=1/3$ being the boundary of the stable
solid phase. With the melting of the solid, the correlation changes to
local diffused triangular symmetry that is characteristic of the
hexatic phase.  In contrast, in the hexatic phase, the correlation
function is short ranged and exhibits an exponential decay.

\subsubsection{Solid Order Parameter}

The structure factor, denoted as
$\langle \psi_{\mathbf{q}} \rangle \equiv (1/N)\langle
\rho_{\mathbf{q}} \rho_{-\mathbf{q}} \rangle$, can effectively
differentiate between various phases, such as solid, hexatic, liquid,
and modulated liquid phases.  In the solid phase,
$\langle \psi_{\mathbf{q}} \rangle$ exhibits a distinct six-fold
symmetry pattern with peaks located at two positions:
$\mathbf{G}_1 = (0, \pm 2 \pi / a_y)$ and
$\mathbf{G}_2 = (\pm 2\pi / a, \pm \pi / a_y)$, corresponding to the
underlying triangular lattice structure. As we transition to the
hexatic phase, these six intensity maxima become broader while
remaining on a constant radius circle at $q = 2\pi / a$. In an
equilibrium melting transition, the value of the solid order parameter
at the transition point is $\langle \psi^*_{\mathbf{G}_2}\rangle\approx 0.4$,
which we take as a cutoff value for the solid--hexatic melting
transition. In a simple liquid with spherical symmetry, this
broadening continues until the peaks overlap, forming a characteristic
ring pattern.

It's worth noting that the presence of an external periodic potential
in this context introduces explicit symmetry breaking by imposing
density modulations, leading to the condition
$\langle \psi_{\mathbf{G}_1} \rangle > \langle \psi_{\mathbf{G}_2}
\rangle$.The other four quasi Bragg peaks located at $\mathbf{G}_2$,
especially at higher ratcheting frequencies, indicate the emergence of
quasi-long-ranged positional order (QLRO). Thus, we use their mean
value as the order parameter for the solid phase, denoted as
$\langle \psi_{\mathbf{G}_2} \rangle$.

\section*{Identifying the orientational order}
\subsubsection{Hexatic order and correlation}
The local orientational order is quantified using the complex order parameter
$\psi_6^{i}$ defined as
\begin{equation}
  \label{eq:local_psi6}
  \psi_6^{i}=\frac{1}{n}\sum_{k=1}^n e^{\complexi 6 \theta_{ik}},
\end{equation}
where $n$ is the number of topological neighbors of the $i^{\rm th}$
particle, identified using Voronoi tessellation. The angle
$\theta_{ik}$ is the angle made by the separation vector
$\vec{r}_{ik}$ with the $x$-axis. The global orientational order
parameter is then defined as
\begin{equation}
  \label{eq:global_psi6}
   \left \langle \psi_6 \right \rangle=\left \langle \left |\frac{1}{N} \sum_{i=1}^N
     \psi_6^{i} \right |^2\right \rangle
\end{equation}

Hexatic correlation is defined by
$g_6(r)=\langle {\psi_6^i}^*\psi_6^j \delta(r-r_{ij})\rangle$,where
${\psi_6^i}^*$ denotes the complex conjugate of $\psi_6^i$ , the local
bond orientational order and $r_{ij}$ is the separation between
particles $i$ and $j$. In the solid phase, it is anticipated that the
function $g_6(r)$ will remain constant with respect to distance $r$
due to the presence of long-range hexatic order. For a single hexatic
with quasi-long-range order (QLRO), it is expected that $g_6(r)$ will
follow a power-law decay characterized by $g_6(r)\sim
r^{-\eta_6}$.
According to the KTHNY theory, $\eta_6$ is anticipated
to converge to the value of $1/4$ as the hexatic melting point is
approached from lower temperatures. Following the melting, it is
expected that the function $g_6(r)$ will exhibit an exponential decay.
The external potential induces a density modulation along the
direction of the drive and consequently the orientational order in the
system is maintained.

\end{appendices}

\end{document}